\documentclass{article}
\usepackage{times}
\usepackage{amssymb}
\usepackage[latin1]{inputenc}
\usepackage{proof}

\usepackage{epsfig}
\usepackage{latexsym}

\usepackage[dvips]{color}

\usepackage{url}

\def\int{\mathbb{Z}}

\usepackage{float}
\floatstyle{ruled}
\newfloat{algo}{h}{lop}
\floatname{algo}{Algorithm}

\def\proof{{\bf Proof: }}
\def\endproof{}
\def\qed{\hfill$\Box$\vskip0.8em}

\newtheorem{theorem}{Theorem}
\newtheorem{lemma}{Lemma}
\newtheorem{example}{Example}

\newcounter{countlem}

\def\savelemmacounter{\setcounter{countlem}{\value{lemma}}}
\def\restorelemmacounter{\setcounter{lemma}{\value{countlem}}}
\def\savethcounter{\setcounter{countlem}{\value{theorem}}}
\def\restorethcounter{\setcounter{theorem}{\value{countlem}}}
\def\namelem#1{\setcounter{#1}{\value{lemma}}}
\def\putlem#1{\setcounter{lemma}{\value{#1}}}
\def\nameth#1{\setcounter{#1}{\value{theorem}}}
\def\putth#1{\setcounter{theorem}{\value{#1}}}

\newcommand\ForAuthors[1]
 {\par\smallskip                     
  \begin{center}
   \fbox
   {\parbox{0.9\linewidth}
    {\raggedright\sc--- #1}
   }
  \end{center}
  \par\smallskip                     %
 }

\def\onevar{\mathcal{V}_1}
\def\flat{\mathcal{F}}
\def\c{\mathcal C}
\def\onevarh{\mathcal{V}_1Horn}
\def\flath{\mathcal{F}Horn}
\def\ch{\mathcal CHorn}
\def\reach{\mathsf{reach}}
\def\known{\mathsf{known}}
\def\empcl{\Box}

\def\atge{\ge}
\def\atlt{<}
\def\atltss{<_s}

\def\sublt{\prec}

\def\abs{\sqsubseteq}
\def\abss{\abs}
\def\absh{\abs_h}
\def\pmodels{\vDash_{\mathrm{p}}}
\def\pnmodels{\nvDash_{\mathrm{p}}}

\def\repl{\rightarrow}
\def\replres{\rightarrow_{\mathcal R}}
\def\splres{\rightarrow_{spl}}
\def\ssplres{\rightarrow_{nspl}}
\def\predssplres{\rightarrow_{\sspreds-nspl}}

\def\atordfsplres{\Rightarrow_{\atlt,\phi}}
\def\subordfsplres{\Rightarrow_{\sublt,\phi}}

\def\subordselfssplreplres{\Rrightarrow_{\sublt_s,\phi,\mathcal R}}
\def\subordselnosplreplres{\Rrightarrow_{\sublt_s,\phi_0,\mathcal R}}
\def\subordselfssplreplresh{\Rrightarrow_{\sublt_s,\phi_h,\mathcal R}}
\def\atordselfssplreplres{\Rrightarrow_{\atlt_s,\phi,\mathcal R}}
\def\atordres{\Rightarrow_{\atlt}}
\def\atordselres{\Rrightarrow_{\atlt_s}}
\def\absres{\blacktriangleright}
\def\absresh{{\blacktriangleright_h}}

\def\inst{\mathsf{I}}
\def\insti{\mathsf{I}_i}
\def\instone{\mathsf{I}_1}
\def\insttwo{\mathsf{I}_2}
\def\instthree{\mathsf{I}_3}
\def\instfour{\mathsf{I}_4}

\def\proj{\pi}
\def\comp{\mathsf{comp}}
\def\eps{\mathsf{eps}}
\def\ov{\mathsf{ov}}
\def\ground{\mathsf{gr}}
\def\fv{\mathsf{fv}}
\def\sspreds{\mathcal Q}
\def\epspreds{\mathcal Q_1}
\def\grpreds{\mathcal Q_0}

\def\ssp#1{\overline{#1}}
\def\qrep#1{\underline{#1}}
\def\varset{{\bf X}}
\def\varone{{\bf x}_1}
\def\vartwo{{\bf x}_2}
\def\varthree{{\bf x}_3}
\def\varr{{\bf x}_r}
\def\varn{{\bf x}_n}
\def\varrone{{\bf x}_{r+1}}
\def\varrtwo{{\bf x}_{r+2}}

\def\varrr{{\bf x}_{2r}}

\def\varione{{\bf x}_{i_1}}

\def\varik{{\bf x}_{i_k}}

\def\varlione{{\bf x}_{i^l_1}}
\def\varlikl{{\bf x}_{i^l_{k_l}}}

\def\rvarset{{\bf X}_r}

\def\ngrr{\mathsf{Ngrr}}
\def\ngrone{\mathsf{Ngr_1}}
\def\ngr{\mathsf{Ngr}}
\def\ngs{\mathsf{Ngs}}
\def\ng{\mathsf{Ng}}
\def\gone{\mathsf{G_1}}
\def\g{\mathsf{G}}
\def\u{\mathsf{U}}

\begin{document}
\title{Flat and One-Variable Clauses: Complexity of Verifying
Cryptographic Protocols with Single Blind Copying}
\author{Helmut Seidl \quad Kumar Neeraj Verma\\
Institut f\"ur Informatik, TU M\"unchen, Germany\\
{\tt \{seidl,verma\}@in.tum.de}}

\maketitle

\begin{abstract}
Cryptographic protocols with single blind copying were defined and
modeled by Comon and Cortier using the new class $\mathcal C$ of
first order clauses.  They showed its
satisfiability problem to be in 3-DEXPTIME.  We improve this result by
showing that satisfiability for this class is NEXPTIME-complete, using
new resolution techniques.  We show satisfiability to be DEXPTIME-complete
if clauses are Horn, which is what is required for modeling cryptographic
protocols. While translation to Horn clauses only gives a DEXPTIME upper
bound for the secrecy problem for these protocols, we further show that
this secrecy problem is actually DEXPTIME-complete.
\end{abstract}


\section{Introduction}
\label{sec:intro}

Several researchers have pursued modeling of cryptographic protocols using
first order clauses~\cite{Blanchet:prolog,comon:rta03,Weidenbach:crypto} 
and related formalisms like
tree automata and set constraints\cite{cc:tamem,glrv:jlap04,Monniaux:SAS99}.
While protocol insecurity is NP-complete in case of a bounded number of sessions
\cite{Turuani:np}, this is helpful only for detecting some attacks.
For certifying protocols, the number of sessions cannot be bounded, although
we may use other safe abstractions.
The approach using first order clauses
is particularly useful for this class of problems.
A common safe abstraction is to allow a bounded number of nonces, i.e.
random numbers, to be used in infinitely many sessions.
Security however
still remains undecidable~\cite{cc:tamem}. Hence further restrictions are 
necessary to obtain decidability. 

In this direction, Comon and Cortier~\cite{comon:rta03,cortier:thesis}
proposed the notion of protocols with single blind copying. 
Intuitively this restriction means that
agents are allowed to copy at most one piece of data blindly in any
protocol step, a restriction satisfied by most protocols in the literature. 
Comon and Cortier modeled the secrecy problem for these protocols using 
the new class $\c$ of first order clauses, 
and showed satisfiability for $\c$ to be 
decidable~\cite{comon:rta03} in 3-DEXPTIME~\cite{cortier:thesis}. 
The NEXPTIME lower bound is easy. We show in this paper
that satisfiability of this class is in NEXPTIME, thus NEXPTIME-complete. 
If clauses
are restricted to be Horn, which suffices for modeling of cryptographic
protocols,  we show that
satisfiability is DEXPTIME-complete (again the lower bound is easy).
While translation to clauses only gives a DEXPTIME upper bound for the
secrecy problem for this class of protocols, 
we further show that the secrecy problem for these protocols 
is also DEXPTIME-complete. 

For proving our upper bounds, we introduce several variants of
standard ordered resolution with selection and splitting~\cite{BG:HAR}.
Notably we consider resolution as consisting of instantiation of clauses, and
of generation of propositional implications. This is in the style 
of Ganzinger and Korovin~\cite{ganzinger:instantiation}, 
but we adopt a slightly different approach, and
generate interesting implications to obtain optimal complexity. 
More precisely, while the approach of~\cite{ganzinger:instantiation}, 
emphasizes a single phase of instantiation
followed by propositional satisfiability checking, we  interleave
generation of interesting instantiations and propositional implications in an
appropriate manner to obtain optimal complexity. 
We further show how this technique
can be employed also in presence of rules for replacement 
of literals in clauses,
which obey some ordering constraints. To deal with the notion of single blind
copying we show  how terms containing a single variable can be decomposed
into simple terms whose unifiers are of very simple forms. As byproducts,
we obtain optimal complexity for several subclasses of  $\c$, 
involving so called {\em flat} and {\em one-variable} clauses.

\noindent
{\bf Outline:}
We start in Section~\ref{sec:fol} by recalling basic notions about first order
logic and resolution refinements. In Section~\ref{sec:protocols} we introduce 
cryptographic protocols with single blind copying, discuss their
modeling using the class $\c$ of first order clauses, and show that their
secrecy problem is DEXPTIME-hard. To decide the class $\c$ we gradually
introduce our techniques by obtaining DEXPTIME-completeness and
NEXPTIME-completeness for one-variables
clauses and flat clauses in Sections~\ref{sec:onevar} and~\ref{sec:flat}
respectively.  In Section~\ref{sec:nh}, the
techniques from the two cases are combined with further ideas to show
that satisfiability for $\c$ is NEXPTIME-complete. In Section~\ref{sec:horn}
we adapt this proof to show that satisfiability for the Horn fragment of 
$\c$ is DEXPTIME-complete.

\section{Resolution}
\label{sec:fol}

We recall standard notions from first order logic.
Fix a signature $\Sigma$ of function symbols each with a given arity, and 
containing at least one zero-ary symbol. 
Let $r$ be the maximal arity of function symbols in $\Sigma$.
Fix a set $\varset = \{\varone,\vartwo,\varthree,\ldots\}$ of variables.
Note that $\varone,\vartwo,\ldots$ (in bold face) are the actual elements of
$\varset$, where as $x,y,z,x_1,y_1,\ldots$ are used to represent arbitrary
elements of $\varset$. The set $T_{\Sigma}(\varset)$ of terms built from
$\Sigma$ and $\varset$ is defined as usual. 
$T_{\Sigma}$ is the set of {\em ground terms}, i.e. those not containing any 
variables. {\em Atoms} $A$ are of the form $P(t_1,\ldots,t_n)$ where $P$ is an 
$n$-ary predicate and $t_i$'s are terms. {\em Literals} $L$ are either positive 
literals $+ A$ (or simply $A$) or negative literals $- A$, where $A$ is an atom.
$-(-A)$ is another
notation for $A$.  $\pm$ denotes $+$ or $-$ and $\mp$ denotes the
opposite sign (and similarly for notations $\pm',\mp',\ldots$).
A {\em clause} is a finite set of literals.
A {\em negative clause} is one which contains only negative literals.
If $M$ is any term, literal or clause then
the set $\fv(M)$ of variables occurring in them is defined as usual.
If $C_1$ and $C_2$ are clauses then $C_1\lor C_2$ denotes $C_1\cup C_2$.
$C \lor \{L\}$ is written as $C \lor L$ (In this notation, we allow the
possibility of $L\in C$). If $C_1,\ldots,C_n$ are clauses such that
$\fv(C_i)\cap \fv(C_j)=\emptyset$ for $i\neq j$, and if $C_i$ is non-empty
for $i\ge 2$, then the clause $C_1 \lor \ldots\lor C_n$ is also written as 
$C_1 \sqcup \ldots \sqcup C_n$ to emphasize this property.
 {\em Ground literals and clauses} are ones not 
containing variables.
A term, literal or clause is {\em trivial} if it contains no
function symbols.
A substitution is a function $\sigma : \varset \rightarrow T_{\Sigma}(\varset)$.
{\em Ground substitutions} map every variable to a ground term.
We write $\sigma = \{x_1\mapsto t_1,\ldots,x_n\mapsto t_n\}$ to say that
$x_i\sigma = t_i$ for $1\le i\le n$ and $x\sigma = x$ for 
$x\notin\{x_1,\ldots,x_n\}$.
If $M$ is a term, literal, clause, substitution or set of such objects,
then the effect $M\sigma$ of
applying $\sigma$ to $M$ is defined as usual.
{\em Renamings} are bijections $\sigma : \varset \rightarrow \varset$. 
If $M$ is a term, literal, clause or substitution, 
then a renaming of $M$ is of the form
$M\sigma$ for some renaming $\sigma$, and an
instance of $M$ is of the form
$M\sigma$ for some substitution $\sigma$.
If $M$ and $N$ are terms or literals then a {\em unifier} of $M$ and $N$
is a substitution $\sigma$ such that $M\sigma = N\sigma$. 
If such a unifier exists then there is also a {\em most general unifier (mgu)},
i.e. a unifier $\sigma$ such that for every unifier $\sigma'$ of $M$ and $N$,
there is some $\sigma''$ such that $\sigma' = \sigma\sigma''$. Most
general unifiers are unique upto renaming: if $\sigma_1$ and $\sigma_2$ are
two mgus of $M$ and $N$ then $\sigma_1$ is a renaming of $\sigma_2$. Hence
we may use the notation $mgu(M,N)$ to denote one of them.
We write $M[x_1,\ldots,x_n]$ to say that $\fv(M) \subseteq \{x_1,\ldots,x_n\}$.
If $t_1,\ldots,t_n$ are terms then $M[t_1,\ldots,t_n]$ denotes 
$M \{x_1 \mapsto t_1, \ldots, x_n \mapsto t_n\}$.
If $N$ is a set of terms them $M[N] = \{M[t_1,\ldots,t_n] \mid
t_1,\ldots,t_n\in N\}$. If $M$ is a set of terms, atoms, literals or clauses
them $M[N] = \bigcup_{m\in M} m[N]$. A {\em Herbrand interpretation} 
$\mathcal H$ is a set of ground atoms. A clause $C$ is {\em satisfied} in 
${\mathcal H}$ if for 
every ground substitution $\sigma$, either $A \in\mathcal H$ for
some  $A \in C\sigma$, or 
$A \notin {\mathcal H}$ for some $-A \in C\sigma$. 
A set $S$ of clauses is satisfied in $\mathcal H$ if every clause of $S$
is satisfied in $\mathcal H$.  If such a $\mathcal H$
exists then $S$ is {\em satisfiable}, and $\mathcal H$ is a {\em Herbrand model}
of $S$.
A {\em Horn clause} is one containing at most one positive literal. If a set
of Horn clauses is satisfiable then it has a least Herbrand model
wrt the subset ordering.

Resolution and its refinements are well known methods for
 testing satisfiability of clauses.
Given a strict partial order $\atlt$ on atoms, a literal
$\pm A$ is {\em maximal} in a clause $C$ if there is no literal
$\pm' B\in C$ with $A \atlt B$.
{\em Binary ordered resolution} and {\em ordered factorization} 
wrt ordering $\atlt$ are defined by the following two rules respectively: 

\[
\prooftree
C_1\lor A \quad   - B \lor C_2
\justifies C_1\sigma \lor C_2\sigma
\endprooftree 
\quad\quad\quad\quad\prooftree
C_1 \lor \pm A \lor \pm B
\justifies C_1\sigma\lor A\sigma
\endprooftree
\]

\noindent
where $\sigma = mgu(A,B)$ in both rules,
$A$ and $B$ are maximal in the left and right premises 
respectively of the first rule, and
$A$ and $B$ are both maximal in the premise of the second rule.
We rename the premises of the first rule before resolution so that
they don't share variables.  The ordering
$\atlt$ is {\em stable} if: whenever $A_1 \atlt A_2$ then 
$A_1\sigma \atlt A_2\sigma$ for all substitutions $\sigma$.  We
write $S \atordres S \cup \{C\}$
 to say that $C$ is obtained by one application of
the binary ordered resolution or binary factorization rule on clauses in
$S$ (the subscript denotes the ordering used). 

Another resolution rule is 
{\em splitting}. This can be described using {\em tableaux}. A {\em tableau}
is of the form $S_1 \mid \ldots \mid S_n$, 
where $n\ge 0$ and each $S_i$, called a {\em branch} of the tableau,
 is a set of 
clauses (the $\mid$ operator is associative and commutative). 
A tableau is {\em satisfiable} if at least one of its branches is satisfiable.
The tableau
is called {\em closed} if each $S_i$ contains the empty clause, denoted
$\empcl$. The {\em splitting}
step on tableaux is defined by the rule $${\mathcal T} \mid S \splres
{\mathcal T} \mid (S \setminus \{C_1\sqcup C_2\}) \cup \{C_1\} \mid
(S \setminus \{C_1\sqcup C_2\}) \cup \{C_2\}$$ whenever $C_1\sqcup C_2 \in S$
and $C_1$ and $C_2$ are non-empty. $C_1$ and $C_2$ are called {\em components}
 of the clause $C_1 \sqcup C_2$ being split.
It is well known that splitting preserves satisfiability of tableaux.
We may choose to apply splitting eagerly,
or lazily or in some other fashion. Hence we define a {\em splitting strategy}
to be a function $\phi$ such that $\mathcal T \splres \phi(\mathcal T)$ for all
tableaux $\mathcal T$. 
The relation $\atordres$ is extended to tableaux as expected.
Ordered resolution with splitting strategy is then
defined by the rule $${\mathcal T}_1 \atordfsplres \phi({\mathcal T}_2)
\textrm{ whenever } {\mathcal T}_1 \atordres {\mathcal T}_2$$ 
This provides us with a well known
sound and complete method for testing satisfiability.
For any binary relation $R$,
$R^*$ denotes the reflexive transitive closure of $R$, and $R^+$ 
denotes the transitive closure of $R$. 
\begin{lemma}
\label{lemma:ordspl:compl}
For any set $S$ of clauses, for any stable ordering $\atlt$, and for
any splitting strategy $\phi$, $S$ is unsatisfiable iff
$S \atordfsplres^* \mathcal T$ for some closed $\mathcal T$.
\end{lemma}

If all predicates are zero-ary then the resulting clauses are
{\em propositional clauses}. In this case we write $S \pmodels T$ to say
that every Herbrand model of $S$ is a Herbrand model of $T$. This notation
will also be used when $S$ and $T$ are sets of first order clauses, by treating
every (ground or non-ground) atom as a zero-ary predicate.
For example $\{P(a),- P(a)\} \pmodels \empcl$ but $\{P(x),-P(a)\}\pnmodels
\empcl$.  $S\pmodels \{C\}$ is also written as $S\pmodels C$. If
$S\pmodels C$ then clearly
$S\sigma \pmodels C\sigma$ for all substitution $\sigma$.

\section{Cryptographic Protocols}
\label{sec:protocols}

We assume that $\Sigma$ contains the binary functions $\{\_\}_{\_}$ and
$\langle\_,\_\rangle$ 
denoting encryption and pairing. {\em Messages} are terms of
$T_{\Sigma}(\varset)$.
A {\em state} is of the form $S(M_1,\ldots,M_n)$ where $S$ with arity $n$
is from a finite set of {\em control points}
and $M_i$ are messages. It denotes an agent at control point $S$
with messages $M_i$ in its memory.  An {\em initialization state} is a
state not containing variables. We assume some strict partial order $<$ on the
set of control points. A {\em protocol rule} is of the form
$$S_1(M_1,\ldots,M_m):\mathsf{recv}(M) \rightarrow 
S_2(N_1,\ldots,N_n):\mathsf{send}(N)$$ where $S_1 < S_2$, $M_i,N_j$ are messages,
and $M$ and $N$ are each either a message, or a dummy symbol $?$ indicating
nothing is received (resp. sent). For secrecy analysis we can replace $?$
by some public message, i.e. one which is known to everyone including the
adversary.
The rule says that an agent in state
$S_1(M_1,\ldots,M_m)$ can receive message $M$, send a message $N$, and then
move to state $S_2(N_1,\ldots,N_n)$, thus also modifying the messages in
its memory. A {\em protocol} is a finite set of initialization states
and protocol rules. This model is in the style of~\cite{durgin:fmsp99} and
\cite{cc:tamem}. The assumption of single blind copying then says that
each protocol rule contains at most one variable (which may occur anywhere
any number of times in that rule). For
example, the public-key Needham-Schroeder protocol 

\[\begin{array}{r l}
\hspace*{1pt}
A \rightarrow B : & \{A,N_A\}_{K_B}\\
\hspace*{1pt}
B \rightarrow A : & \{N_A,N_B\}_{K_B}\\
\hspace*{1pt}
A \rightarrow B : & \{N_B\}_{K_B}
\end{array}\]

\noindent
is written in our notation as follows. 
For every pair of agents $A$ and $B$
in our system (finitely many of
 them suffice for finding all attacks
against secrecy~\cite{comon:esop03,comon:rta03}) we have two nonces
$N^1_{AB}$ and $N^2_{AB}$ to be used in sessions where $A$ plays the
initiator's role and $B$ plays the responder's role.
We have initialization states 
$\mathsf{Init_0}(A,N^1_{AB})$ and $\mathsf{Resp_0}(B,N^2_{AB})$ for all
agents $A$ and $B$.
Corresponding to the three lines in the protocol
we have rules for all agents $A$ and $B$

\[\begin{array}{r@{:}l@{\rightarrow}r@{:}l}
\mathsf{Init_0}(A,N^1_{AB}) &  \mathsf{recv}(?)  &  
\mathsf{Init_1}(A,N^1_{AB}) & \mathsf{send} (\{\langle A,N^1_{AB}
\rangle\}_{K_B})\\
\mathsf{Resp_0}(B,N^2_{AB}) &  \mathsf{recv}(\{\langle A,x\rangle\}_{K_B})
 &  \mathsf{Resp_1}(B,x,N^2_{AB}) &  \mathsf{send}(\{\langle x,N^2_{AB}\rangle\}_{K_A})\\
\mathsf{Init_1}(A,N^1_{AB})  &  \mathsf{recv}(\{\langle N^1_{AB},x\rangle\}_{K_A})
 & \mathsf{Init_2}(A,N^1_{AB},x)  &  \mathsf{send}(\{x\}_{K_B})\\
\mathsf{Resp_1}(B,x,N^2_{AB}) & \mathsf{recv}(\{N^2_{AB}\}_{K_B})  & 
\mathsf{Resp_2}(B,x,N^2_{AB})  &  \mathsf{send}(?) 
\end{array}\]

Any initialization state can be created any number of times and any 
protocol rule can be executed any number of times. The adversary has
full control over the network: all messages received by agents are
actually sent by the adversary and all messages sent by agents are
actually received by the adversary. The adversary can obtain
new messages from messages he knows, e.g. by performing encryption
and decryption.  To model this using Horn
clauses, we create a unary predicate $\reach$ to model reachable
states, and a unary predicate $\known$ to model messages known to the
adversary. The initialization state $S(M_1,\ldots,M_n)$
is then modeled by the clause
$\reach(S(M_1,\ldots,M_n))$, where $S$ is a new function symbol we create.
The protocol rule 
$$S_1(M_1,\ldots,M_m):\mathsf{recv}(M) \rightarrow
S_2(N_1,\ldots,N_n):\mathsf{send}(N)$$
is modeled by the clauses
$$\begin{array}{c}\known(N) \lor - \reach(S_1(M_1,\ldots,M_m)) \allowbreak
\lor \allowbreak - \known(M) \\
\reach(S_2(N_1,\ldots,N_n)) \allowbreak \lor \allowbreak - \reach(\allowbreak S_1(\allowbreak M_1,\allowbreak \ldots,\allowbreak M_m)) \allowbreak \lor \allowbreak -
\known(M)\end{array}$$
Under the assumption
of single blind copying it is clear that all these clauses are {\em 
one-variable clauses}, i.e. clauses containing at most one variable. 
 We need further clauses to express adversary capabilities. 
The clauses 
$$\begin{array}{c}
\known(\{\varone\}_{\vartwo}) \lor - \known(\varone)\lor - \known(\vartwo)\\
\known(\varone) \lor - \known(\{\varone\}_{\vartwo}) \lor - \known(\vartwo)
\end{array}$$
 express the encryption
and decryption abilities of the adversary. We have similar clauses for
his pairing and unpairing abilities, as well as clauses
$$\known(\allowbreak f\allowbreak (\allowbreak \varone\allowbreak ,\allowbreak \ldots\allowbreak ,\allowbreak \varn\allowbreak )\allowbreak ) \lor - \known(\varone)\lor \ldots \lor -
\known(\varn)$$
 for any function $f$ that the adversary knows to apply. All these
are clearly {\em flat clauses}, i.e. clauses of the form
$$C=\bigvee_{i=1}^k \pm_i P_i(f_i(x^i_1,\ldots,x^i_{n_i})) \lor
\bigvee_{j=1}^l \pm_j Q_j(x_j)$$
where $\{x^i_1,\ldots,x^i_{n_i}\} = \fv(C)$ for $1\le i\le k$.
Asymmetric keys, i.e. keys $K$ such that message $\{M\}_K$ can only be decrypted
with the inverse key $K^{-1}$,  are also easily dealt with using flat and 
one-variable clauses.
The adversary's knowledge of other data $c$
like agent's names, public keys, etc are expressed by clauses $\known(c)$.
Then the least Herbrand model of this set of clauses describes exactly the
reachable states and the messages known to the adversary. Then to check whether
some message $M$ remains secret, we add the clause
$ - \known(M)$ and check whether the resulting set is satisfiable.

A set of clauses is in the class $\onevar$ if each of its members is a
one-variable clause. 
A set of clauses is in the class $\flat$ if each of its members is a
flat clause. More generally we have the class 
$\c$ proposed by Comon and Cortier~\cite{comon:rta03,cortier:thesis}: 
a set of clauses $S$ is in the class $\c$ if for each $C\in S$ one of the
following conditions is satisfied.
\begin{enumerate}
\item $C$ is a one-variable clause
\item $C=\bigvee_{i=1}^k \pm_i P_i(u_i[f_i(x^i_1,\ldots,x^i_{n_i})]) \lor
\bigvee_{j=1}^l \pm_j Q_j(x_j)$,
where for  $1\le i\le k$ we have $\{x^i_1,\ldots,x^i_{n_i}\} = \fv(C)$ and
$u_i$ contains at most one variable.
\end{enumerate}
If all clauses are Horn then we have the corresponding classes
$\onevarh$, $\flath$ and $\ch$. 
Clearly the classes $\onevar$ (resp. $\onevarh$) and $\flat$ (resp. 
$\flath$) are included in the class $\c$ (resp.
$\ch$) since the $u_i$'s above can be trivial.
Conversely any clause set in $\c$ can be considered as containing just
flat and one-variable clauses. This is because we can replace a clause
$C \lor \pm P(u[f(x_1,\ldots,x_n)])$ by the clause
$C \lor \pm Pu(f(x_1,\ldots,x_n))$ and add clauses
$ - Pu(x) \lor P(u[x])$ and $Pu(x) \lor  - P(u[x])$ where $Pu$ is a
fresh predicate. This transformation takes polynomial time and preserves
satisfiability of the clause set.  Hence now we 
need to deal with just flat and one-variable clauses.
In the rest of the paper we derive optimal complexity results
for all these classes.

Still this  only gives us an upper bound for the secrecy problem of 
protocols since the clauses could be more general than necessary.
It turns out, however, that this is not the case. In order to show this
we rely on a reduction of 
the reachability problem for {\em alternating pushdown systems (APDS)}. 
In form of Horn clauses, an {\em APDS} is a finite 
set of clauses of the form 
\begin{enumerate}
\item[(i)] $P(a)$ where $a$ is a zero-ary symbol
\item[(ii)] $P(s[x])\lor - Q(t[x])$ where $s$ and $t$ involve only unary function 
symbols, and
\item[(iii)] $P(x) \lor - P_1(x) \lor - P_2(x)$
\end{enumerate}
 Given any set $S$ of {\em definite}
clauses (i.e. Horn clauses having some positive literal),
a ground atom $A$ is {\em reachable} if $A$ is in the least
Herbrand model of $S$, i.e. if $S \cup
\{ - A\}$ is unsatisfiable. Reachability in APDS is
DEXPTIME-hard~\cite{kozen:alternation}. 
We encode this problem into secrecy of protocols, 
as in~\cite{durgin:fmsp99}. 
Let $K$ be a (symmetric) key not known to the adversary.
Encode atoms $P(t)$ as messages $\{\langle P,t\rangle\}_K$, by treating
$P$ as some data. 
Create initialization states $S_1$ and $S_2$ 
(no message is stored in the states).
Clauses (i-iii) above are translated as 
$$\begin{array}{l l l}
S_1: & \mathsf{recv}(?) & \rightarrow S_2:\mathsf{send}(\{\langle P,a\rangle\}_K)\\
S_1: & \mathsf{recv}(\{\langle Q,t[x]\rangle\}_K) \allowbreak & \rightarrow \allowbreak S_2:\mathsf{send}(\{\langle P,s[x]\rangle\}_K)\\
S_1: & \mathsf{recv}\allowbreak (\allowbreak \langle \allowbreak \{\allowbreak \langle \allowbreak P_1\allowbreak ,\allowbreak x\allowbreak \rangle\allowbreak \}_K,\{\allowbreak \langle \allowbreak P_2\allowbreak ,\allowbreak x\allowbreak \rangle\}_K\rangle) 
& \rightarrow S_2:\mathsf{send}(\{\langle P,x\rangle\}_K)
\end{array}$$
The intuition is that the adversary cannot decrypt messages encrypted with
$K$. He also cannot encrypt messages with $K$. He can only forward messages
which are encrypted with $K$. However he has the ability to pair messages.
This is utilized in the translation of clause (iii). Then a message
$\{M\}_K$ is known to the adversary iff $M$ is of the form $\langle P,t\rangle$ 
and $P(t)$ is reachable in the APDS.
\begin{theorem}
\label{th:hardness}
Secrecy problem for cryptographic protocols with single blind copying,
with bounded number of nonces but unbounded number of sessions is 
DEXPTIME-hard, even if no message is allowed to be stored at any control point.
\end{theorem} 

\section{One Variable Clauses: Decomposition of Terms}
\label{sec:onevar}

We first show that satisfiability for the classes $\onevar$ and $\onevarh$ is 
DEXPTIME-complete.  We recall also that although we consider only unary 
predicates, this is no restriction in the case of one-variable clauses, 
since we can encode atoms $P(t_1,\ldots,t_n)$ as
$P'(f_n(t_1\ldots,t_n))$ for fresh $P'$ and $f_n$ for every $P$ of arity $n$.
As shown in~\cite{comon:rta03,cortier:thesis}, ordered
resolution on one-variable clauses, for a suitable ordering,
 leads to a linear bound on the height of
terms produced. This does not suffice for obtaining a DEXPTIME upper bound 
and we need to examine the forms of unifiers
produced during resolution. We consider terms containing at most one variable
(call them {\em one-variable terms}) to be compositions of
simpler terms.
A non-ground one-variable term $t[x]$ is called {\em reduced} if it is not of 
the form $u[v[x]]$ for any non-ground non-trivial one-variable terms $u[x]$ 
and $v[x]$. The term $f(g(x),h(g(x)))$ for example
is not reduced because it can be written
as $f(x,h(x))[g(x)]$. The term $f'(x,g(x),a)$ is reduced. Unifying it with the
reduced term $f'(h(y),g(h(a)),y)$ produces ground unifier $\{x\mapsto h(y)[a],
y\mapsto a\}$ and both $h(y)$ and $a$ are strict subterms of the given terms.
Indeed we find:

\newcounter{lemsxtyred}
\namelem{lemsxtyred}
\def\lemsxtyredstatement{
Let $s[x]$ and $t[y]$ be reduced, non-ground and non-trivial terms
where $x\neq y$ and $s[x] \neq t[x]$. 
If $s$ and $t$ have a unifier $\sigma$ then
$x\sigma, y\sigma \in U[V]$ where $U$ is the set of non-ground
(possibly trivial) strict subterms of $s$ and $t$, and $V$ is the set of ground 
strict subterms of $s$ and $t$.
}
\begin{lemma}
\label{lemma:sxty:red}
\lemsxtyredstatement
\end{lemma}
\proof
See Appendix~\ref{appone}.

In case both terms (even if not reduced) have the same variable
we have the following easy result: 
\newcounter{lemsxtx}
\namelem{lemsxtx}
\def\lemsxtxstatement{
Let $\sigma$ be a unifier of two non-trivial,
non-ground and distinct one-variable terms $s[x]$ and $t[x]$. Then 
$x\sigma$ is a ground strict subterm of $s$ or of $t$.
}
\begin{lemma}
\label{lemma:sxtx}
\lemsxtxstatement
\end{lemma}
\proof
See Appendix~\ref{appone}.

In the following one-variable clauses are simplified to
involve only reduced terms.
\newcounter{lemdecomp}
\namelem{lemdecomp}
\def\lemdecompstatement{
Any non-ground one-variable term $t[x]$ can be uniquely written as
$t[x] = t_1[t_2[\ldots[t_n[x]]\ldots]]$ where $n\ge 0$ and each $t_i[x]$ is
non-trivial, non-ground and reduced. This decomposition
can be computed in time polynomial in the size of $t$.
}
\begin{lemma}
\label{lemma:decomp}
\lemdecompstatement
\end{lemma}
\proof 
We represent $t[x]$ as a DAG by doing maximal sharing of subterms.
If $t[x]=x$ then the result is trivial. Otherwise
let $N$ be the position in this graph, other than the root node,
 closest to the root such that 
$N$ lies on every path from the root to the node corresponding to the
subterm $x$. 
Let $t'$ be the strict subterm of $t$ at position $N$ and
let $t_1$ be the term obtained from $t$ by replacing the sub-DAG at $N$
by $x$. Then $t=t_1[t']$ and $t_1$ is reduced.
We then recursively decompose $t'$.

Uniqueness of decomposition follows from Lemma~\ref{lemma:sxty:red}.\qed
\endproof 

Above  and elsewhere, if $n=0$ then
$t_1[t_2[\ldots[t_n[x]]\ldots]]$ denotes $x$.
Now if a clause set contains a clause 
$C = C' \lor \pm P(t[x])$, with $t[x]$ being non-ground,
if $t[x] = t_1[\ldots[t_n[x]]\ldots]$
where each $t_i$ is non-trivial and reduced, then we 
create fresh predicates $Pt_1\ldots t_i$ for $1\le i\le n-1$ and replace $C$
by the clause $C' \lor \pm Pt_1\ldots t_{n-1}(t_n[x])$. 
Also we add clauses
$Pt_1\ldots t_i(t_{i+1}[x]) \lor - Pt_1\ldots t_{i+1}(x)$ and
$ - Pt_1\ldots t_i(t_{i+1}[x]) \allowbreak \lor \allowbreak Pt_1\ldots t_{i+1}\allowbreak (\allowbreak x\allowbreak )$ 
for $0\le i \le n-2$ to our clause set. 
Note that the predicates $Pt_1\ldots t_i$ are considered invariant under
renaming of terms $t_j$. For $i=0$, $Pt_1\ldots t_i$ is same as $P$. 
Our transformation preserves satisfiability of the clause set. 
By Lemma~\ref{lemma:decomp} this takes polynomial time and eventually all
non-ground literals in clauses are of the form $\pm P(t)$ with reduced $t$.
Next if the clause set is of the form $S\cup\{C_1\cup C_2\}$, 
where $C_1$ is non-empty and has only ground literals, and $C_2$ is non-empty
and has only non-ground literals,
then we do splitting to produce $S\cup\{C_1\}\mid S\cup \{C_2\}$.
This process produces at most exponentially many branches each of which 
has polynomial size. Now it suffices to decide satisfiability of each branch in
DEXPTIME. Hence now we assume that each clause is either:\\
\hspace*{7pt}
({\bf Ca}) a ground clause, or\\
\hspace*{7pt}
({\bf Cb}) a clause containing exactly one variable,
each of whose literals is of the form
      $\pm P(t[x])$ where $t$ is non-ground and reduced.\\
Consider a set  $S$ of clauses of type Ca and Cb. We show how to decide
satisfiability of the set $S$. Wlog we assume that all clauses in $S$ of type
Cb contain the variable $\varone$.
Let $\ng$ be the set of non-ground terms $t[\varone]$ occurring as arguments in 
literals in $S$. 
Let $\ngs$ be the set of non-ground subterms $t[\varone]$ of terms in $\ng$.
We assume that $\ng$ and $\ngs$ always contain the trivial term $\varone$,
otherwise we add this term to both sets.
Let $\g$ be the set of ground subterms of terms occurring as arguments in
literals in $S$. The sizes of $\ng,\ngs$ and $\g$ are polynomial.
Let $S^\dag$ be the set of clauses of type Ca and Cb 
which only contain literals of the form 
$\pm P(t)$ for some $t\in \ng\cup \ng[\ngs[\g]]$ (observe that $\g \subseteq
\ngs[\g] \subseteq \ng[\ngs[\g]]$). The size of $S^\dag$ is at most exponential.

For resolution we use ordering $\sublt$: $P(s) \sublt Q(t)$ iff $s$ is a
strict subterm of $t$. We call $\sublt$ the subterm ordering without
causing confusion. This is clearly stable. 
This is the ordering that we are going to use throughout this
paper. 
In particular this means that if a clause contains literals $\pm P(x)$ and
$\pm' Q (t)$ where $t$ is non-trivial and contains $x$, then we cannot 
choose the literal $\pm P(x)$ to resolve upon in this clause.
Because of the simple form of unifiers of reduced terms we have:

\newcounter{lemovresol}
\namelem{lemovresol}
\def\lemovresolstatement{
Binary ordered resolution and ordered factorization, wrt the subterm
ordering, on clauses in $S^\dag$ produces clauses which are
again in $S^\dag$ (upto renaming). 
}
\begin{lemma}
\label{lemma:ovresol}
\lemovresolstatement
\end{lemma}
\proof  
Factorization on a ground clause doesn't produce any new clause.
Now suppose we factorize the non-ground clause $C[\varone]
\lor \pm P(s[\varone])\lor \pm P(t[\varone])$ 
to produce the clause $C[\varone]\sigma\lor \pm P(s[\varone])\sigma$ where
$\sigma=mgu(s[\varone],t[\varone])$. If the premise has only trivial literals
then factorization is equivalent to doing nothing. Otherwise by ordering
constraints, $s$ and $t$ are non-trivial.
By Lemma~\ref{lemma:sxtx} either $s[\varone]=t[\varone]$ in which case
factorization does nothing,
or $\varone\sigma$ is a ground subterm of $s[\varone]$ or of $t[\varone]$. 
In the latter case all literals in $(C[\varone]\lor P(s[\varone])\sigma$ 
are of the form $\pm' Q(t'[\varone]\sigma))$ where $t'[\varone]\in \ng$ 
and $\varone\sigma\in \g\subseteq \ngs[\g]$.

Now we consider binary resolution steps.
We have the following cases:
\begin{itemize}
\item If both clauses are ground then the result is clear.

\item Now consider both clauses $C_1[\varone]$ and 
$C_2[\varone]$ to be non-ground. Before resolution we rename
the second clause to obtain $C_2[\vartwo]$.
Clearly all literals in $C_1[\varone]$ and $C_2[\varone]$ are of the form
$\pm Q(u[\varone])$ where $u[\varone]\in \ng$.  
Let $C_1[\varone] = C'_1[\varone] \lor P(s[\varone])$ and 
$C_2[\vartwo] = - P(t[\vartwo]) \lor C'_2[\vartwo]$ 
where $P(s[\varone])$ and $-P(t[\vartwo])$ are the literals
to be resolved upon in the respective clauses.
If $s[\varone]$ and $t[\vartwo]$ are unifiable then from
Lemma~\ref{lemma:sxty:red}, one of the following cases hold:

\begin{itemize}
\item $s[\varone] = \varone$ (the case where $t[\vartwo] = \vartwo$
is treated similarly). From the definition of $\sublt$,
for $P(s[\varone])$ to be chosen for resolution,
all literals in $C'_1[\varone]$ are of the form $\pm Q(\varone)$.
The resolvent is $C[\vartwo] = C'_1[\varone]\sigma \cup C'_2$,
where $\sigma = \{\varone \mapsto t[\vartwo]\}$.
Each literal in $C'_1[\varone] \sigma$ is of the form 
$\pm Q(t[\vartwo])$ and each 
literal in $C'_2[\vartwo]$ is of the form $\pm Q(t'[\vartwo])$ where 
$t'\in \ng$. Hence $C[\varone] \in S^\dag$.

\item $s[\varone] = t[\varone]$. Then the resolvent is $C'_1[\varone]\lor
C'_2[\varone]$. 

\item $s[\varone]$ and $t[\vartwo]$ have a mgu $\sigma$ such that 
      $\varone\sigma, \vartwo\sigma \in \ngs[\g]$.
      The resolvent $C'_1[\varone]\sigma \lor C'_2[\vartwo]\sigma$ 
      has only ground atoms of the form $\pm Q(t')$ where $t'\in \ng[\ngs[\g]]$.
\end{itemize}

\item Now let the first clause $C_1[\varone]=C'_1[\varone]\lor \pm 
P(t[\varone])$ be non-ground, and the second clause $C_2 = \mp P(s) \lor C'_2$ 
be ground with $\pm P(t[\varone])$ and $\mp P(s)$ being the
respective literals chosen from $C_1[\varone]$ and $C_2$ for resolution.
All literals in $C_1[\varone]$ are of the form $\pm' Q(t'[\varone])$ 
with $t'\in \ng$.  All literals in $C_2$ are of the
form $\pm' Q(t')$ with $t'\in \ng[\ngs[\g]]$. 
Suppose that $s$ and $t[\varone]$ do unify.
We have the following cases:
\begin{itemize}
\item $s\in \ngs[\g]$. Then the resolvent $C = 
C'_1[\varone]\sigma \cup C'_2$ where $\sigma = \{\varone\mapsto g\}$ 
where $g$ is subterm of $s$. As $s\in \ngs[\g]$ hence
$g\in \ngs[\g]$. Hence all literals in $C'_1[\varone]\sigma$ are of
the form $\pm Q(t')$ where $t'\in \ng[\ngs[\g]]$. Hence $C \in S^\dag$.

\item
Now suppose $s\in \ng[\ngs[\g]]\setminus \ngs[\g]$. 
We must have $s = s_1[s_2]$ for some non-trivial $s_1[\varone] \in \ng$ and
some $s_2\in \ngs[\g]$.
This is the interesting case which shows why the
terms remain in the required form during resolution. 
The resolvent is
$C = C'_1[\varone]\sigma \lor C'_2$ where $\sigma=\{\varone\mapsto g\}$
is the mgu of $t[\varone]$ and $s$ for some ground term $g$.
As $t[g] = s_1[s_2]$, $\sigma_1 = \{\varone\mapsto g,\vartwo\mapsto s_2\}$ 
is a unifier of the terms $t[\varone]$ and $s_1[\vartwo]$. 
By Lemma~\ref{lemma:sxty:red} we have the following cases:
\begin{itemize}
\item $t[\varone]=\varone$, so that $g=s\in \ng[\ngs[\g]]$. By definition
of $\sublt$, for $\pm P(t[\varone])$ to be chosen for resolution, 
all literals in $C_1[\varone]$ must be of the form $\pm' Q(\varone)$. 
Hence all literals in $C'_1\sigma$ are of the
form $\pm' Q(g)$.  Hence $C \in S^\dag$.

\item $t[\varone] = s_1[\varone]$.
Then $g = s_2\in \ngs[\g]$.
Hence all literals in $C'_1\sigma$ are of the form
$\pm' Q(t'[g])$ where $t'[\varone]\in \ng$. Hence $C \in S^\dag$.

\item $g=\varone\sigma\in \ngs[\g]$. 
Hence all literals in $C'_1\sigma$ are of the form
$\pm' Q(t'[g])$ where $t'\in \ng$. Hence $C \in S^\dag$.
\qed

\end{itemize} 
\end{itemize}
\end{itemize}
\endproof 

Hence to decide satisfiability of $S \subseteq S^\dag$, we keep 
generating new clauses of $S^\dag$ by doing ordered binary resolution and 
ordered factorization wrt the subterm ordering till no new clause 
can be generated, 
and then check whether the empty clause has been produced. Also recall
that APDS consist of Horn one-variable clauses. Hence:
\begin{theorem}
\label{theorem:onevar}
Satisfiability for the classes $\onevar$ and
$\onevarh$ is DEXPTIME-complete.
\end{theorem}

\section{Flat Clauses: Resolution Modulo Propositional Reasoning}
\label{sec:flat}

Next we show how to decide the class $\flat$ of flat clauses in NEXPTIME.
This is well known when the maximal arity $r$ is a constant, or when all
non-trivial literals in a clause have the same {\em sequence}
 (instead of the same 
{\em set}) of variables.  But we are not aware of a proof of NEXPTIME upper
bound in the general case. We show how to obtain 
NEXPTIME upper bound in the general
case, by doing resolution modulo propositional reasoning.
While this constitutes an interesting result of its own,
the techniques allow us  to deal with the full class $\c$ efficiently.
Also this shows that the generality of the class $\c$ does
not cost more in terms of complexity.
An {\em $\epsilon$-block} is a one-variable clause which contains only
trivial literals. A complex clause $C$ is a flat clause
$\bigvee_{i=1}^k \pm_i P_i(f_i(x^i_1,\ldots,x^i_{n_i})) \lor
\bigvee_{j=1}^l \pm_j Q_j(x_j)$ in which $k\ge 1$. Hence a flat clause is either
a complex clause, or an 
{\em $\epsilon$-clause} which is defined to be a disjunction of 
$\epsilon$-blocks, i.e. to be  of the form 
$B_1[x_1] \sqcup \ldots \sqcup  B_n[x_n]$ where each $B_i$ is an 
$\epsilon$-block. $\epsilon$-clauses are difficult to deal with, hence we
split them to produce $\epsilon$-blocks.
Hence define {\em $\epsilon$-splitting} as the restriction of the splitting rule
in which one of the components is an $\epsilon$-block.

Recall that $r$ is the maximal arity of symbols in $\Sigma$. 
Upto renaming, any complex clause $C$ is such that
$\fv(C) \subseteq \rvarset = \{\varone,\ldots,\varr\}$, and 
any $\epsilon$-block $C$ is such that $\fv(C) \subseteq \{\varrone\}$.
The choice of $\varrone$ is not crucial. Now notice that
ordered resolution between complex clauses and $\epsilon$-blocks only 
produces flat
clauses, which can then be split to be left with only complex and
$\epsilon$-blocks. E.g. 
Resolution between $$P_1(\varone) \lor -P_2(\vartwo) 
\lor P_3(f(\varone,\vartwo)) \lor -P_4(g(\vartwo,\varone))$$ and
$$P_4(g(\varone,\varone)) \lor -P_5(h(\varone)) \lor P_6(\varone)$$ produces 
$$P_1(\varone) \lor -P_2(\varone) \lor P_3(f(\varone,\varone))
\lor -P_5(h(\varone)) \lor P_6(\varone)$$ 
Resolution between $$P_2(\varrone) \quad\textrm{ and }\quad
-P_2(f(\varone,\vartwo)) \lor P_3(\varone) \lor P_4(\vartwo)$$ produces
$P_3(\varone) \lor P_4(\vartwo)$ which can then be split. The point is that
we always choose a non-trivial literal from a clause for resolution, if there
is one. As there are finitely many complex clauses and $\epsilon$-blocks this gives
us a decision procedure. Note however that the number of complex clauses
is doubly exponential. This is because we allow clauses of the form
$P_1(f_1(\varone,\varone,\vartwo)) \lor P_2(f_2(\vartwo,\varone)) \lor
P_3(f_3(\vartwo,\varone,\vartwo)) \lor ...$, i.e. the nontrivial terms
contain arbitrary number of repetitions of  variables in arbitrary order.
The number of such variable sequences of $r$ variables is exponentially
many, hence the number of clauses is doubly exponential. Letting
the maximal arity $r$ to be a constant, or forcing all 
non-trivial literals in a clause
to have the same variable sequence would have produced only exponentially
many clauses. In presence of splitting, this would have given us the well-known
NEXPTIME upper bound, which is also optimal. 
But we are not aware of a proof of NEXPTIME upper
bound in the general case. To obtain NEXPTIME upper bound
in the general case we introduce the technique of
resolution modulo propositional reasoning.

For a clause $C$, define the set of its projections as
$\proj(C) = C[\rvarset]$. Essentially projection involves making certain
variables in a clause equal. As we saw, resolution between two complex clauses
amounts to propositional resolution between their projections.
Define the set $\u = \{f(x_1,\ldots,x_n) \mid f \in \Sigma
\textrm{ and each }x_i\in\rvarset\}$ of size exponential in $r$.
Resolution between $\epsilon$-block $C_1$ and a good
complex clause  $C_2$ amounts to
propositional resolution of a clause from $C[\u]$ with $C_2$. Also
note that propositional resolution followed by further projection is 
equivalent to projection followed by propositional resolution.  Each complex
clause has exponentially many projections. This suggests that we can
compute beforehand the exponentially many projections of complex clauses and
exponentially many instantiations of $\epsilon$-blocks. All new complex clauses
generated by propositional resolution are ignored.
But after several such propositional resolution steps, we may get an 
$\epsilon$-clause, which should then be split and instantiated and used for
obtaining further propositional resolvents. 
In other words we only compute such
propositionally implied $\epsilon$-clauses, do splitting and instantiation and
iterate the process. This generates all resolvents upto propositional
implication. 
We now formalize our approach.
We start with the following observation which 
is used in this and further sections.

\begin{lemma}
\label{lemma:sequnif}
Let $x_1,\ldots,x_n,y_1,\ldots,y_n$ be variables, not necessarily distinct,
but with $\{x_1,\ldots,x_n\} \cap \{y_1,\ldots,y_n\} = \emptyset$.
Then the terms $f(x_1,\ldots,x_n)$ and
$f(y_1,\ldots,y_n)$ have an mgu $\sigma$ such that
$\{x_1,\ldots,x_n\}\sigma \subseteq \{x_1,\ldots,x_n\}$
and $y_i\sigma = x_i\sigma$ for $1\le i\le n$.
\end{lemma}

For a set $S$ of clauses, $\comp(S)$ is the set of complex clauses in $S$,
 $\eps(S)$ the set of $\epsilon$-blocks in $S$,
$\proj(S) = \bigcup_{C\in S} \proj(C)$ and
$\inst(S) = \proj(\comp(S)) \cup \eps(S)[\varrone] \cup \eps(S)[\u]$. 
For sets $S$ and $T$ of complex clauses and $\epsilon$-blocks,
$S \abs T$ means that:\\
-- if $C\in S$ is a complex clause then $\inst(T) \pmodels \proj(C)$, and\\
-- if $C\in S$ is an $\epsilon$-block then $C[\varrone] \in \eps(T)[\varrone]$.\\
For tableaux ${\mathcal T}_1$ and ${\mathcal T}_2$ involving only complex clauses and
$\epsilon$-blocks we write ${\mathcal T}_1 \abs {\mathcal T}_2$ if
${\mathcal T}_1$ can be written as $S_1 \mid \ldots \mid S_n$ 
and ${\mathcal T}_2$
can be written as $T_1 \mid \ldots \mid T_n$ (note same $n$)
such that $S_i \abs T_i$ for $1\le i\le n$. Intuitively 
${\mathcal T}_2$ is a succinct representation of ${\mathcal T}_1$.
Define the splitting strategy $\phi$ as the one which repeatedly
applies $\epsilon$-splitting on a tableau as long as possible. The
relation $\subordfsplres$ provides us a sound and complete method for
testing unsatisfiability.
We define the alternative procedure for testing unsatisfiability by using
succinct representations of tableaux.  We define $\absres$ by the rule:
$\mathcal T \mid S \absres \mathcal T \mid S \cup \{B_1\} \mid \ldots \mid
S \cup \{B_k\}$ 
whenever $\inst(S) \pmodels C = B_1[\varione] \sqcup \ldots
\sqcup B_k[\varik]$, $C$ is an $\epsilon$-clause, and
$1\le i_1,\ldots,i_k \le r+1$.  Then $\absres$ simulates $\subordfsplres$:
\newcounter{lemabsimulate}
\namelem{lemabsimulate}
\def\lemabsimulatestatement{
If $S$ is a set of complex clauses and $\epsilon$-blocks, $S \abs T$ and $S
\subordfsplres {\mathcal T}$, then all clauses occurring in ${\mathcal
T}$ are complex clauses or $\epsilon$-blocks and $T \absres^* {\mathcal T}'$
for some ${\mathcal T}'$ such that ${\mathcal T} \abs {\mathcal T}'$.
}
\begin{lemma}
\label{lemma:abs:simulate}
\lemabsimulatestatement
\end{lemma}
\proof 
We have the following ways in which ${\mathcal T}$ is obtained from $S$
by doing one resolution step followed by splitting:

\begin{itemize}

\item
We resolve two $\epsilon$-blocks 
$C_1$ and $C_2$ of $S$ to get an $\epsilon$-block $C$, and
${\mathcal T} = S \cup \{C\}$. Then 
$\{C_1[\varrone],C_2[\varrone]\}\pmodels C[\varrone]$. Also
as $S \abs T$ we have  $\{C_1[\varrone],C_2[\varrone]\} \subseteq 
\eps(T)[\varrone]$.  We have $\inst(T) \pmodels C[\varrone]$.
Hence $T \absres T \cup \{C[\varrone]\}$
and clearly $S \cup \{C\} \abs T \cup \{C\}$.

\item
We resolve an $\epsilon$-block $C_1[\varrone]$ with a complex clause 
$C_2[\varone,\ldots,\varr]$, both from $S$ upto renaming, and we have
$C_1[\varrone]\in \eps(T)[\varrone]$ and $\inst(T) \pmodels 
\proj(C_2)$. By ordering constraints, we have $C_1[\varrone] = 
C'_1[\varrone] \lor \pm P(\varrone)$ and 
$C_2[\varone,\ldots,\varr] = \mp P(f(x_1,\ldots,x_n)) \lor 
C'_2[\varone,\ldots,\varr]$ so that resolution produces
$C[\varone,\ldots,\varr]
 = C'_1[f(x_1,\ldots,x_n)] \lor C'_2[\varone,\ldots,\varr]$.
Clearly $C_1[\u] \cup \{C_2[\varone,\ldots,\varr]\} \pmodels 
C[\varone,\ldots,\varr]$. Also $\proj(C_1[\u]) = C_1[\u]$.
Hence $\inst(T) C_1[\u] \cup \proj(C_2) \pmodels 
\proj(C) \supseteq \{C[\varone,\ldots,\varr]\}$.
\begin{itemize}
\item
If $C'_1$ is not empty or if $C'_2$ has some non-trivial literal then
$C$ is a complex clause and ${\mathcal T} = S \cup \{C\} \abs T$.
\item If $C'_1$ is empty and $C'_2$ has only trivial literals then 
$C[\varone,\ldots,\varr]$ is 
an $\epsilon$-clause of the form $B_1[\varione] \sqcup \ldots \sqcup
B_k[\varik]$ 
with $1\le i_1,\ldots,i_k\le r$.
${\mathcal T} = S \cup \{B_1\} \mid \ldots \mid S \cup \{B_k\}$.
Since $\inst(T) \pmodels C[\varone,\ldots,\varr]$, 
hence $T \absres {\mathcal T}'$ where ${\mathcal T}' = 
T \cup \{B_1\}  \mid \ldots \mid T \cup \{B_k\}$ and we have
${\mathcal T} \abs {\mathcal T}'$.
\end{itemize}

\item
We resolve two complex clauses $C_1[\varone,\ldots,\varr]$ and 
$C_2[\varone,\ldots,\varr]$, both from $S$ upto renaming, and we have
$\inst(T) \pmodels \proj(C_1)$ and $\inst(T) \pmodels \proj(C_2)$.
First we rename the second clause as $C_2[\varrone,\ldots,\varrr]$ by
applying the renaming $\sigma_0 = \{\varone \mapsto \varrone,
\ldots,\varr \mapsto \varrr\}$.
By ordering constraints, $C_1[\varone,\ldots,\varr]$ is of the form 
$C'_1[\varone,\ldots,\varr] \lor 
\pm P(f(x_1,\ldots,x_n))$ and $C_2[\varrone,\ldots,\varrr]$ is of the form
$\mp P(f(y_1,\ldots,y_n)) \lor C'_2[\varrone,\ldots,\varrr]$ so that 
$\pm P(f(x_1,\ldots, x_n))$ and $\mp P(f(y_1,\ldots,y_n))$ are
the literals to be resolved from the respective clauses. 
By Lemma~\ref{lemma:sequnif}, the resolvent is $C = C'_1[\varone,\ldots,\varr]
\sigma \lor C'_2 [\varrone,\ldots,\varrr]\sigma$ where $\sigma$ is such that
$\{x_1,\ldots,x_n\} \sigma \subseteq \{x_1,\ldots,x_n\}$
and $y_i\sigma = x_i\sigma$ for $1\le i\le n$. 
$C$ is obtained by propositional resolution from
$C_1\allowbreak [\allowbreak \varone,\allowbreak \ldots,\allowbreak \varr\allowbreak ]\allowbreak \sigma \allowbreak \in\allowbreak  \proj(C_1)$ 
and $C_2[\varrone,\ldots,\varrr]\sigma
= C_2[\varone,\ldots,\varr]\sigma_0\sigma \in \proj(C_2)$.
Hence $\proj(C_1) \cup \proj(C_2) \pmodels C[\varone,\ldots,\varr]$. 
Hence $\proj(\proj(C_1)) \cup \proj(\proj(C_2)) = \proj(C_1) \cup
\allowbreak \proj\allowbreak (C_2) 
\pmodels \proj(C)$.  As $\inst(T) \pmodels \proj(C_1)$ and
$\inst(T) \pmodels \proj(C_2)$.
hence $\inst(T) \pmodels \proj(C) \supseteq \{C[\varone,\ldots,\varr])\}$.

\begin{itemize}
\item
If either $C'_1$ or $C'_2$ contains a non-trivial literal then 
$C$ is a complex clause and ${\mathcal T} = S \cup \{C\} \abs T$. 
\item
If $C'_1$ and $C'_2$ contain only trivial literals then 
$C[\varone,\ldots,\varr]$ is an
$\epsilon$-clause of the form $B_1[\varione]\sqcup\ldots\sqcup B_k[\varik]$
with $1\le i_1, \ldots, i_k \le r$.
${\mathcal T} = S \cup \{B_1\} \mid \ldots \mid S \cup \{B_k\}$.
As $\inst(T) \pmodels C[\varone,\ldots,\varr]$ we have
$T \absres \mathcal T'$ where $\mathcal T' =
T \cup \{B_1\} \mid \ldots \mid T \cup \{B_k\}$.
Also $\mathcal T \abs \mathcal T'$.
\end{itemize}

\item
$C[\varone,\ldots,\varr]$ is a renaming of a 
complex clause in $S$, and we factor $C\allowbreak [\allowbreak \varone\allowbreak ,\allowbreak \ldots\allowbreak ,\allowbreak \varr\allowbreak ]$ 
to get a complex clause $C[\varone,\ldots,\varr]\sigma$ where
$\rvarset\sigma\subseteq \rvarset$, and 
${\mathcal T} = S\cup\{C[\varone,\ldots,\varr]\sigma\}$.
$C[\varone,\ldots,\varr]\sigma \in \proj(C)$.
Hence $\proj\allowbreak (\allowbreak \{\allowbreak C\allowbreak [\allowbreak \varone\allowbreak ,\allowbreak \ldots,\varr\allowbreak ] \sigma\}) \subseteq 
\proj(\proj(C)) = \proj(C)$.
As $S \abs T$ hence $\inst(T) \pmodels \proj(C)$.
Hence $\inst(T) \pmodels \proj(\{C[\varone,\allowbreak \ldots,\allowbreak \varr]\sigma\})$.
Hence we have $\mathcal T = S \cup \{C[\varone,\allowbreak \ldots,\allowbreak \varr]\sigma\} \abs T$.\qed
\end{itemize}
\endproof 
Hence we have completeness of $\absres$:
\begin{lemma}
\label{lemma:abs:complete}
If a set $S$ of good complex clauses and $\epsilon$-blocks is unsatisfiable then 
$S \absres^* \mathcal T$ for some closed $\mathcal T$.
\end{lemma}
\proof 
By Lemma~\ref{lemma:ordspl:compl},
$S \subordfsplres^* S_1 \mid \ldots \mid S_n$ such
that each $S_i\owns \empcl$.
As $S \abs S$, hence by Lemma~\ref{lemma:abs:simulate}, we have some
$T_1,\ldots,T_n$ such that $S\absres^* T_1\mid\ldots\mid T_n$ and $S_i\abs T_i$ for
$1\le i\le n$. Since $\empcl \in S_i$ and $\empcl$ is an $\epsilon$-block,
hence $\empcl\in T_i$ for $1\le i\le n$.
\qed
\endproof 

Call a set $S$ of complex clauses and $\epsilon$-blocks
{\em saturated} if the following condition is satisfied:
if $\inst(S) \pmodels B_1[\varione] \sqcup \ldots \sqcup B_k[\varik]$ 
with $1\le i_1,\ldots,i_k\le r+1$, each $B_i$ being an $\epsilon$-block, 
then there is some $1\le j\le k$ such that $B_j[\varrone] \in S[\varrone]$.

\begin{lemma}
\label{lemma:abs:saturate}
If $S$ is a satisfiable set of complex clauses and $\epsilon$-blocks 
then $S \absres^* \mathcal T \mid T$ for some $\mathcal T$ and some
saturated set $T$ of complex clauses and $\epsilon$-blocks,
such that $\empcl \notin T$.
\end{lemma}
\proof 
We construct a sequence $S = S_0 \subseteq S_1\subseteq S_2 \subseteq \ldots$
of complex clauses and $\epsilon$-blocks such that $S_i$ is satisfiable and
$S_i \absres^* S_{i+1} \mid \mathcal T_i$ for some $\mathcal T_i$ for each $i$. 
$S=S_0$ is satisfiable by assumption.
Now assume we have already defined $S_0,\ldots,S_i$ and $\mathcal T_0,
\ldots,\mathcal T_{i-1}$.
Let $C^l = B^l_1[\varlione] \sqcup \ldots \sqcup B^l_k[\varlikl]$
for $1\le l\le N$ be all the possible $\epsilon$-clauses such that 
$\inst(S_i) \pmodels C^l$, $1\le i^l_1,\ldots,i^l_{k_l}\le r+1$.
Since $S_i$ is satisfiable,
 $S_i \cup \{C^l \mid 1\le l\le N\}$ is satisfiable.
Since $\varlione,\ldots,\varlikl$ are mutually distinct for $1\le l\le N$,  
there are $1\le j_l\le k_l$ for $1\le l\le N$ such that
$S_i \cup \{B^l_{j_l}\mid 1\le l\le N\}$ is satisfiable.
Let $S_{i+1} = S_i \cup \{B^l_{j_l}\mid 1\le l\le N\}$. 
$S_{i+1}$ is satisfiable.
Also it is clear that $S_i \absres^* S_{i+1} \mid \mathcal T_i$ for some
$\mathcal T_i$.
If $S_{i+1} = S_i$ then $S_i$ is saturated, otherwise $S_{i+1}$ has
strictly more $\epsilon$-blocks upto renaming. 
As there are only finitely many $\epsilon$-blocks upto renaming, 
eventually we will end up with a saturated set $T$
in this way. Since $T$ is satisfiable, $\empcl \notin T$. From construction
it is clear that there is some $\mathcal T$ such that $S \absres^* \mathcal T
\mid T$.
 \qed
\endproof 

\begin{theorem}
\label{theorem:flatsat}
Satisfiability for the class $\flat$ is NEXPTIME-complete.
\end{theorem}
\proof 
The lower bound comes from reduction of satisfiability of positive set
constraints which is NEXPTIME-complete~\cite{akvw:set}. For the upper bound
let $S$ be a finite set of flat clauses. 
Repeatedly apply $\epsilon$-splitting to obtain 
$f(S)=S_1 \mid \ldots \mid S_m$.  $S$ is satisfiable iff
some $S_i$ is satisfiable. 
The number $m$ of branches in $f(S)$ is at most exponential. 
Also each branch has size linear in the size of $S$. 
We non-deterministically choose some $S_i$ and check its satisfiability
in NEXPTIME.

Hence wlog we may assume that the given set $S$ has only complex clauses and
$\epsilon$-blocks.  We non-deterministically choose a certain number of
$\epsilon$-blocks $B_1,\ldots,B_N$ 
and check that $T =  S_1 \cup \{B_1,\ldots,B_N\}$
is saturated and $\empcl\notin T$.
By Lemma~\ref{lemma:abs:saturate}, 
if $S$ is satisfiable then clearly there is such a set $T$.
Conversely if there is such a set $T$, then whenever $T \absres^* 
\mathcal T$, we will have $\mathcal T = T \mid \mathcal T'$ 
for some ${\mathcal T}'$. Hence we can never have $T \absres^* {\mathcal T}$
where $\mathcal T$ is closed.  Then by Lemma~\ref{lemma:abs:complete}
we conclude that $T$ is satisfiable. Hence $S \subseteq T$ is also satisfiable.

Guessing the set $T$ requires non-deterministically 
choosing from among exponentially many $\epsilon$-blocks.
To check that $T$ is saturated, for every  
$\epsilon$-clause $C = B_1[\varione] \sqcup \ldots \sqcup B_k[\varik]$, 
with $1\le i_1, \ldots, i_k\le r+1$, and $B_j[\varrone]
 \notin T[\varrone]$ for $1\le j\le k$, we check that
$\inst(T) \pnmodels C$, i.e. 
$\inst(T) \cup \lnot C$ is propositionally satisfiable
(where $\lnot(L_1\lor\ldots\lor L_n)$ denotes $\{-L_1,\ldots,-L_n\}$).
This can be checked in NEXPTIME since
propositional satisfiability can be checked in NPTIME.
We need to do such checks for at most exponentially many possible
values of $C$. \qed
\endproof

\section{Combination: Ordered Literal Replacement}
\label{sec:nh}

Combining flat and one-variable clauses creates additional difficulties.
First observe that resolving a one variable clause 
$C_1 \lor \pm P(f(s_1[x],\ldots,s_n[x]))$
with a complex clause $\mp P(f(x_1,\ldots,x_n)) \lor C_2$ produces a 
one-variable clause. If $s_i[x]=s_j[x]$ for all $x_i=x_j$, and if
$C_2$ contains a literal $P(x_i)$ then the resolvent contains a literal
$P(s_i[x])$. The problem now is that even if
$f(s_1[x],\ldots,s_n[x])$ is reduced, $s_i[x]$ may not be reduced. E.g.
$f(g(h(x)),x)$ is reduced but $g(h(x))$ is not reduced. Like in
Section~\ref{sec:onevar} we may think of replacing this literal by simpler
literals involving fresh predicates. Firstly we have to ensure that in this
process we do not generate infinitely many predicates. Secondly
it is not clear that mixing ordered resolution steps
with replacement of literals is still complete. Correctness is easy to show
since the new clause is in some sense equivalent to the old deleted clause.
However deletion of clauses arbitrarily can violate completeness of the 
resolution procedure. The key factor which preserves completeness is that
we replace literals by smaller literals wrt the given ordering $\atlt$.

Formally a {\em replacement rule} is of  the form 
$A_1 \repl A_2$ where $A_1$ and $A_2$
are (not necessarily ground) atoms.  The clause set 
{\em associated} with this rule is $\{A_1 \lor - A_2, 
 - A_1 \lor A_2\}$.
Intuitively such a replacement rule says that  $A_1$ and
$A_2$ are equivalent. The clause set $cl(\mathcal R)$
associated with a set $\mathcal R$
of replacement rules is the union of the clause sets associated with the 
individual replacement rules in $\mathcal R$.
Given a stable ordering $\atlt$ on atoms, a 
replacement rule $A_1 \repl A_2$ is {\em ordered} iff
$A_2\atlt A_1$. 
We define the relation $\repl_{\mathcal R}$ as: $S \repl_{\mathcal R}
(S \setminus \{\pm A_1 \sigma \lor C \}) \cup \{\pm A_2\sigma \lor C\}$ 
whenever $S$ is a set of clauses, $\pm A_1\sigma \lor C \in S$, 
$A_1 \repl A_2 \in \mathcal R$ and $\sigma$ is some substitution. 
Hence we replace literals in a clause by smaller literals.
The relation is extended to tableaux as usual.
This is reminiscent of the well-studied  
case of resolution with some equational theory
on terms. There, however, the ordering
$\atlt$ used for resolution is compatible with the equational theory and one
essentially works with the equivalence classes of terms and atoms. This is not
the case here.

Next note that in the above resolution example, even if 
$f(s_1[x],\ldots,s_n[x])$ is non-ground, some $s_i$ may be ground. 
Hence the resolvent may have ground as well as non-ground literals. We avoided
this in Section~\ref{sec:onevar} by initial preprocessing. Now we may think
of splitting these resolvents during the resolution procedure. This however
will be difficult to simulate using the alternative resolution procedure
on succinct representations of tableaux because we will generate doubly 
exponentially many one-variable clauses. To avoid this
we use a variant of splitting 
called {\em splitting-with-naming}~\cite{voronkov:splitting}. 
Instead of creating two branches after
splitting, this rule puts both components into the same set, but with
tags to simulate branches produced by ordinary splitting. 
Fix a finite set $\mathbb P$ of predicate symbols. 
$\mathbb P$-clauses are clauses whose predicates are all from $\mathbb P$.
Introduce fresh zero-ary predicates 
$\ssp{C}$ for $\mathbb P$-clauses $C$ modulo renaming, i.e. 
$\ssp{C_1}=\ssp{C_2}$
iff $C_1\sigma=C_2$ for some renaming $\sigma$. Literals $\pm \ssp{C}$
for $\mathbb P$-clauses $C$ are  {\em splitting literals}.
The {\em splitting-with-naming} rule is defined as:
$S \ssplres (S \setminus \{C_1 \sqcup C_2\}) \cup \{C_1 \lor -\ssp{C_2},
\ssp{C_2} \lor C_2\}$ where $C_1 \sqcup C_2\in S$,
$C_2$ is non-empty and has only non-splitting literals, and
$C_1$ has at least one non-splitting literal. 
Intuitively $\ssp{C_2}$ represents the negation of $C_2$.
We will use both
splitting and splitting-with-naming according to some predefined strategy.
Hence for a finite set
$\sspreds$ of splitting atoms, define {\em $\sspreds$-splitting}
as the restriction of the splitting-with-naming rule where the splitting
atom produced is restricted to be from $\sspreds$. Call this restricted
relation as $\predssplres$. This is extended to tableaux as usual.
Now once we have generated the clauses
$C_1 \lor -\ssp{C_2}$ and $\ssp{C_2} \lor C_2$ we would like to keep resolving
on the second part of the second clause till we are left with the clause
$\ssp{C_2}$ (possibly with other positive splitting literals)
 which would then be resolved with the first clause to produce
$C_1$ (possibly with other positive splitting literals)
 and only then the literals in $C_1$ would be resolved upon.
Such a strategy cannot be ensured by ordered resolution, hence we introduce
a new rule.
An ordering $\atlt$ over non-splitting atoms is extended to the ordering
$\atltss$ by letting $q\atltss A$ whenever $q$ is a splitting atom
and $A$ is a non-splitting atom, and $A \atltss B$
whenever $A,B$ are non-splitting atoms and $A \atlt B$.
We define {\em modified ordered binary resolution} by the following rule:\\
\strut\hfill$\prooftree
C_1\lor A \quad   - B \lor C_2
\justifies C_1\sigma \lor C_2\sigma
\endprooftree$\hfill\ 

\noindent
where $\sigma = mgu(A,B)$ and the following conditions are satisfied:\\
(1) $C_1$ has no negative splitting literal, and $A$ is maximal in $C_1$.\\
(2) (a) either $B\in \sspreds$, or\\
\hspace*{15pt}(b) $C_2$ has no negative splitting literal, and $B$ is maximal in $C_2$.\\
As usual we rename the premises before resolution so that 
they don't share variables. This rule says that we must select a negative
splitting literal to resolve upon in any clause, provided the clause has
at least one such literal. If no such literal is present in the clause, then
the ordering $\atltss$  enforces that a positive splitting literal will
not be selected as long as the clause has some non-splitting literal.
We write 
$S \atordselres S \cup \{C\}$ to say that $C$ is obtained by one application of
the modified binary ordered resolution or the (unmodified)
ordered factorization rule on clauses in $S$. This is extended to tableaux
as usual.
A {\em $\sspreds$-splitting-replacement strategy} is  a 
function $\phi$ such that $\mathcal T (\predssplres \cup\splres \cup\replres)^* 
\phi(\mathcal T)$  for any tableaux $\mathcal T$. Hence we allow both normal
splitting and $\sspreds$-splitting.
Modified ordered resolution with $\sspreds$-splitting-replacement
strategy $\phi$ is defined by the relation: $S \atordselfssplreplres \phi(T)$ 
whenever $S \cup cl(\mathcal R) \atordselres T$. 
This is extended to tableaux as usual.  The above modified ordered
binary resolution rule can be considered as an instance of 
{\em ordered resolution with selection}~\cite{BG:HAR}, which is known 
to be sound and complete even with splitting and its variants. 
Our manner of extending $\atlt$ to $\atltss$ is essential for completeness.
We now show 
that soundness and completeness hold even under arbitrary ordered replacement 
strategies. It is not clear to the authors if such rules have been studied elsewhere.
 Wlog we forbid the useless case of replacement rules
containing splitting symbols.
The relation $\atlt$ is  {\em enumerable} if 
the set of all 
ground atoms can be enumerated as $A_1,A_2,\ldots$ 
such that if $A_i \atlt A_j$ then $i < j$. 
The subterm ordering is enumerable.

\newcounter{thcompl}
\nameth{thcompl}
\def\thcomplstatement{
Modified ordered resolution, wrt a stable and enumerable ordering,
with splitting and $\sspreds$-splitting and ordered  literal
replacement is sound and complete for any strategy. I.e.
for any set $S$ of $\mathbb P$-clauses, for any strict stable and enumerable
partial order $\atlt$ on atoms, for any set
$\mathcal R$ of ordered replacement rules,
for any finite set $\sspreds$ of splitting atoms, and for any 
$\sspreds$-splitting-replacement strategy $\phi$, $S \cup cl(\mathcal  R)$
is unsatisfiable iff $S \atordselfssplreplres^* \mathcal T$ 
for some closed $\mathcal T$.
}
\begin{theorem}
\label{th:compl}
\thcomplstatement
\end{theorem}
\proof
See Appendix~\ref{apptwo}.

For the rest of this section fix a set $\mathbb S$ of one-variable 
$\mathbb P$-clauses and complex $\mathbb P$-clauses whose satisfiability 
we need to decide.  Let $\ng$ be the set of non-ground
terms occurring as arguments in literals in the one-variable clauses
of $\mathbb S$. We rename all terms in $\ng$ to contain only the variable
$\varrone$.
Wlog assume $\varrone \in \ng$.
Let $\ngs$ be the set of non-ground subterms of terms in $\ng$,
and $\ngr = \{s[\varrone]\mid s \textrm{ is non-ground and reduced,}
 \allowbreak \textrm{and for some } t, s[t]\in \ngs\}$.
Define $\ngrr = 
\{s_1[\allowbreak \ldots\allowbreak [\allowbreak s_m\allowbreak ]\allowbreak \ldots\allowbreak ]\allowbreak \mid s_1[\ldots[s_n]\ldots] \in \ngs,\allowbreak m\le n,
\textrm{ and each } s_i \allowbreak 
\textrm{ is \allowbreak non-trivial \allowbreak and \allowbreak reduced}\}$.
Define the set of predicates $\mathbb Q = \{Ps \mid P\in\mathbb P,s \in \ngrr\}$.
Note that $\mathbb P \subseteq \mathbb Q$.
Define the set of replacement rules $\mathcal R = 
\{Ps_1\ldots s_{m-1} (s_m[\varrone]) \repl \allowbreak Ps_1\ldots s_m \allowbreak (\allowbreak [\allowbreak \varrone\allowbreak ]\allowbreak ) 
\mid Ps_1\ldots s_m  \in \mathbb Q\}$. They are clearly ordered wrt $\sublt$.
Let $\g$ be the set of ground subterms of terms occurring as arguments in 
literals in $\mathbb S$. 
Define the set $\grpreds = \{\ssp{\pm P(t)} \mid P \in \mathbb P,
t\in \g\}$ of splitting atoms. Their purpose is to remove ground literals from 
a non-ground clause.  All sets defined above have polynomial size. 
Let $\sspreds\supseteq\grpreds$ be any set of splitting atoms. For dealing with
the class $\c$ we only need $\sspreds=\grpreds$, but for a more precise analysis
of the Horn fragment in the next Section, we need $\sspreds$ to also contain 
some other splitting atoms.
We also need the set $\ngrone = \{\varrone\} \cup 
\{f (s_1,\ldots,s_n) \mid \exists g(t_1,\ldots,t_m) \in \ngr \cdot
\{s_1,\ldots,s_n\}=\{t_1,\ldots,t_m\}\}$ which has exponential size.
These terms are produced by resolution of non-ground one-variable
clauses with complex clauses, and are also reduced. 
In the ground case we have the set
$\gone = \{f(s_1,\ldots,s_n) \mid \exists g(t_1,\ldots,t_m)
\in \g \mid \{s_1,\ldots,s_n\}=\{t_1,\ldots,t_m\}\}$ of exponential size.
For a set $\mathbb P'$ of predicates and a set $U$ of terms,
the set $\mathbb P'[U]$ of atoms is defined as usual. For a set $V$ of
atoms the set $-V$ and $\pm V$ of literals is defined as usual.
The following types of clauses will be required during resolution:
\begin{enumerate}
\item[(C1)] clauses $C \lor D$, where
$C$ is an $\epsilon$-block with predicates from $\mathbb Q$, and
$D\subseteq\pm\sspreds$.
\item[(C2)] clauses $C \lor D$  where
$C$ is a renaming of a one-variable clause with literals from 
$\pm \mathbb Q(\ngrone)$,
$C$ has at least one non-trivial literal, and $D\subseteq\pm \sspreds$.
\item[(C3)] clauses $C \lor D$ 
where $C$ is a non-empty clause with literals from 
$\pm \mathbb Q(\ngrone[\ngrr[\gone]])$, and $D\subseteq\pm\sspreds$.
\item[(C4)] clauses $C\lor D$ where $C= 
\bigvee_{i=1}^k \pm_i P_i(f_i(x^i_1,\ldots,x^i_{n_i})) \lor
\bigvee_{j=1}^l \pm_j Q_j(x_j)$ is a complex clause
with each $P_i\in \mathbb Q$, each $n_i\ge2$, each $Q_j \in \mathbb P$ and 
$D\subseteq\pm\sspreds$
\end{enumerate}

We have already argued why we need splitting literals in the above clauses,
and why we need $\ngrone$ instead of $\ngr$ in type C2.
In type C3 we have $\ngrr$ in place of the set $\ngs$ that we had in
Section~\ref{sec:onevar}, to take care of interactions between one-variable 
clauses and complex clauses.  In type C4 the trivial literals involve
predicates only from $\mathbb P$ (and not $\mathbb Q$). This is what ensures 
that we need only finitely many fresh predicates (those from $\mathbb Q
\setminus \mathbb P$) because these are the literals that are involved in
replacements when this clause is resolved with a one-variable clause.
We have also required that each $n_i\ge 2$. This is only to ensure that types
C2 and C4 are disjoint. The clauses that are excluded because of this condition
are necessarily of type C2.

The $\grpreds$-splitting steps that we use in this section consist of
replacing a tableau $\mathcal T \mid S$ by the tableau $\mathcal T \mid
(S \setminus \{C \lor L\}) \cup \{C \lor -\ssp{L},
\ssp{L} \lor L\}$, where $C$ is non-ground, $L\in\pm\mathbb P(\g)$
and $C \lor L \in S$.
The replacement steps we are going to use are of the following kind:\\
(1) replacing clause $C_1[x] = C\lor\pm P(t_1[\ldots[t_n[s[x]]]\ldots])$
by clause $C_2[x] = C \lor \pm Pt_1\ldots t_n(s[x])\}$ where $P\in\mathbb P$,
$s[\varrone]\in\ngr$ is  non-trivial, and $t_1[\ldots[t_n]\ldots] \in \ngrr$.
We have $\{C_1[\varrone]\} \cup cl(\mathcal R)[\ngrr] \pmodels C_2[\varrone]$.
\\
(2) replacing ground clause $C_1 = C\lor\pm P(t_1[\ldots[t_n[g]]\ldots])$
by clause $C_2 = C \lor \pm Pt_1\ldots t_n[g]\}$ where $P\in\mathbb P,
g\in \ngrr[\gone]$ and $t_1[\ldots[t_n]\ldots] \in \ngrr$. This replacement
is done only when $t_1[\ldots[t_n[g]]\ldots] \in \ngrr[\ngrr[\gone]]\setminus
\ngrone[\ngrr[\gone]]$. We have
$\{C_1\} \cup cl(\mathcal R)[\ngrr[\ngrr[\gone]]] \pmodels C_2$.\\
Define the $\grpreds$-splitting-replacement strategy $\phi$ as one which 
repeatedly applies first $\epsilon$-splitting,
then the above $\grpreds$-splitting steps, then the above two 
replacement steps till no further change is possible. Then
$\subordselfssplreplres$ gives us a sound and complete method for
testing unsatisfiability. 

As in Section~\ref{sec:flat} we now define
a succinct representation of tableaux and an alternative resolution procedure
for them. As we said, a literal 
$\ssp{L} \in\grpreds$ represents $-L$. Hence for a clause $C$ we define
$\qrep{C}$ as the clause obtained by replacing every $\pm\ssp{L}$ by the
literal $\mp L$.  This is extended to sets of clauses as usual. Observe that
if $S \pmodels C$ then $\qrep S \pmodels \qrep C$.
As before  $\u = \{f(x_1,\ldots,x_n) \mid
f\in\Sigma, \textrm{ and each } x_i \in \rvarset\}$.
The functions $\eps$ and $\comp$ of Section~\ref{sec:flat} are now modified
to return clauses of type C1 and C2 respectively. 
For a set $S$ of clauses, define $\ov(S)$ as the set of clauses of type C2
in $S$. The function $\proj$ is as before.
We need to define which kinds of instantiations are to be used to generate
propositional implications.
For a clause $C$, define 
\[\begin{array}{r l}
\instone(C) = &  C[\u[\ngrr \cup \ngrr[\ngrr[\gone]]]] 
\cup C[\ngrone] \cup C[\ngrone[\ngrr[\gone]]]\\
\insttwo(C) = & \{C[\varrone]\} \cup C[\ngrr[\gone]]\\
\instthree(C) = & \{C\}\\
\instfour(C) = & \proj(C) \cup C[\ngrr \cup \ngrr[\ngrr[\gone]]]
\end{array}\]

The instantiations defined by $\inst_i$ are necessary for clauses of
type C$i$. Observe that $C[U]\subseteq\instone(C)$.
For a set $S$ of clauses, define $\insti(S) = \bigcup_{C\in S} \insti(C)$. 
For a set $S$ of clauses of type C1-C4 define
$\inst(S) =  \instone(\qrep{\eps(S)}) \cup 
\insttwo(\qrep{\ov(S)}) \cup  \instthree(\qrep{\ground(S)}) \cup
\instfour(\qrep{\comp(S)}) \cup cl(\mathcal R)[\ngrr \cup \ngrr[\ngrr[\gone]]]$.
Note that instantiations of clauses in $cl(\mathcal R)$ are necessary for
the replacement rules, as argued above.
For a set $T$ of clauses define the following properties:
\begin{itemize}
\item
$C$ satisfies property P1$_T$ iff $C[\varrone] \in T$.
\item
$C$ satisfies property P2$_T$ iff $\inst(T) \pmodels \insttwo(\qrep{C})$.
\item
$C$ satisfies property P3$_T$ iff $\inst(T) \pmodels \instthree(\qrep{C})$.
\item
$C$ satisfies property P4$_T$ iff $\inst(T)\pmodels\instfour(\qrep{C})$.
\end{itemize}
For sets of clauses $S$ and $T$, define $S \abss T$ to mean that 
every $C\in S$ is of type C$i$ and satisfies property P$i_T$ for some
$1\le i\le 4$. This is extended to tableaux as usual.
We first consider the effect of one step of the above resolution procedure 
without splitting. Accordingly let $\phi_0$ be the variant of 
$\phi$ which applies 
replacement rules and $\grpreds$-splitting, but no $\epsilon$-splitting. 

\begin{lemma}
\label{lemma:nhresol}
Let $S$ be a set of clauses of type C1-C4. If
$S\subordselnosplreplres S'$ then one of the following 
statements holds.
\begin{itemize}
\item $S' \abss S$
\item $S' = S \cup \{C\} \cup S''$, $C$ is a renaming of
$B_1[\varione]\sqcup \ldots \sqcup
B_k[\varik]\sqcup D$, each $B_i$ is an $\epsilon$-block, 
$1\le i_1,\ldots,i_k\le r$, $D \subseteq\pm\sspreds$,
$\inst(S) \pmodels \qrep C$, and $S''$ is a set of clauses of type C3 and
$\emptyset \pmodels \qrep{S''}$. If $k\ge 2$ then $D$ has no literals $-q$
with $q\in\sspreds\setminus\grpreds$.
\end{itemize}
\end{lemma}
\proof 
The set $S''$ in the second statement will contain the clauses $\ssp{L} \lor
L$ added by $\grpreds$-splitting, while $C$ will be the clause produced by
binary resolution or factoring, possibly followed by applications of
replacement rules and by
replacement of ground literals $L$ by $-\ssp{L}$. Hence $S''=\emptyset$ in all
cases except when we need to perform $\grpreds$-splitting.

First we consider resolution steps where splitting literals are resolved upon.
A positive splitting literal cannot be chosen to resolve upon in a clause
unless the clause
has no literals other than positive splitting literals. Hence this clause
is $C_1 = q \lor q_1\lor\ldots\lor q_m$ of type C1,
The other clause must be $C_2= C'_2 \lor -q$ 
of type C$i$ for some $1\le i\le4$. Resolution produces clause 
$C = C'_2 \lor  q_1\lor\ldots\lor  q_m$ of type C$i$, 
and no replacement or splitting rules apply. 
We have $\{C_1,C_2\}\pmodels C$ and $\{\qrep{C_1},\qrep{C_2}\}\pmodels
\qrep{C}$. Hence $\inst(S) \supseteq \qrep{C_1} \cup \insti(\qrep{C_2}) \pmodels
\insti(\qrep{C})$.
If $i=1$ then the second statement of the lemma holds because 
$\insti(\qrep C)$ contains a renaming of $\qrep C$. If $i>1$ then the first
statement holds.

Now we consider binary resolution steps where no splitting literals are 
resolved upon.
This is possible only when no negative splitting literals are present in
the premises. Then the resolvent has no negative splitting literals. 
$\grpreds$ splitting may create negative splitting literals, but none of them
are from $\sspreds\setminus\grpreds$. Hence the last part of the second
statement of the lemma is always true.
In the following $D,D_1,\ldots$ denote subsets of $\grpreds$.
When we write $C\lor D$, it is implicit that $C$ has no splitting literals.
We have the following cases:

\begin{enumerate}
\item We do resolution between two clauses $C_1$ and $C_2$ from $S$, 
both of type C1, and the resolvent  $C$ is of type C1.
Hence no splitting or replacement rules apply,
${S'} = S \cup \{C\}$, $\inst(S) \supseteq 
\{\qrep{C_1}[\varrone],\qrep{C_2}[\varrone]\} \pmodels \qrep{C}[\varrone]$.
Hence the second statement holds.

\item We do resolution between a clause $C_1[\varrone] = C'_1[\varrone]
\lor D_1 \lor \pm P(\varrone)$, of type C1,
and a clause $C_2[\varrone] = \mp P(t[\varrone]) 
\lor C'_2[\varrone] \lor D_2$, of type C2,
both from $S$ upto renaming, and the resolvent 
is $C[\varrone] = C'_1[t[\varrone]] \lor C'_2[\varrone]\lor  D_1 \lor D_2$.
By ordering constraints $t[\varrone]\in \ngrone$  is non-trivial.
All literals in $C'_1[t[\varrone]]\lor C'_2[\varrone]$ 
are of the form $\pm' Q(t'[\varrone])$  with $t'[\varrone]\in\ngrone$. 
Hence no splitting or replacement rules apply and $S' = S \cup \{C\}$. 
$\qrep{C_1}[\ngrone] \cup \{\qrep{C_2}[\varrone]\}\pmodels \qrep{C}[\varrone]$. 
Hence $\inst(S) \supseteq \instone(\qrep{C_1}) \cup \insttwo(\qrep{C_2}) 
\supseteq
\qrep{C_1}[\ngrone] \cup \qrep{C_2}[\allowbreak \ngrone[\allowbreak \ngrr[\allowbreak \gone]]] \cup
\{\qrep{C_2}[\varrone]\} \cup \qrep{C_2}[\ngrr[\gone]] \pmodels 
\{\qrep{C}[\varrone]\} \cup \qrep{C}[\ngrr[\gone]] =
\insttwo(\qrep{C}[\varrone])$. 
If $C'_1$ is non-empty or $C'_2$ has some 
non-trivial literal
 then $C[\varrone]$ is  of type C2, $S' \abss S$ and the first statement
holds.
If $C'_1$ is empty and $C'_2$ has only trivial
literals, then $C$ is of type C1 and the second statement holds.

\item We do resolution between a clause $C_1[\varrone] = 
C'_1[\varrone] \lor D_1 \lor \pm P(\varrone)$ of type C1,
and a clause $C_2 = \mp P(t) \lor C'_2 \lor D_2$ of type C3,
both from $S$ upto renaming, and the resolvent 
is $C = C'_1[t] \lor C'_2\lor  D_1 \lor D_2$.
We know that $t\in \ngrone[\ngrr[\gone]]$. 
Hence no splitting or replacement rules apply, and
${S'} = S \cup \{C\}$.
$\{C_1[t],C_2\} \pmodels C$.  Hence
$\inst(S)\supseteq \instone(\qrep{C_1}[\varrone]) \cup \instthree(\qrep{C_2}) 
\supseteq \qrep{C_1}[\ngrone[\ngrr[\gone]]] \cup \{\qrep{C_2}\}  \pmodels 
\instthree(\qrep{C}) = \{\qrep{C}\}$.
If $C'_1$ or $C'_2$ is non-empty.
then $C[\varrone]$ is  of type C3, $S' \abss S$ and the first statement 
holds.  If $C'_1$ and $C'_2$ are empty
then $C$ is of type C1 and the second statement holds.

\item We do resolution between a clause $C_1[\varrone] = C'_1[\varrone]
\lor D_1 \lor \pm P(\varrone)$ of type
C1, and a clause $C_2[\varone,\ldots,\varr] = \mp P(x_1,\ldots,x_n) \lor 
C'_2[\varone,\ldots,\varr] \lor D_2$  of type C4, both from $S$ upto renaming, 
and the resolvent is $C[\varone,\ldots,\varr]
 = C'_1[f(x_1,\ldots,x_n)] \lor C'_2[\varone,\ldots,\varr]\lor  D_1 \lor D_2$.
(By ordering constraints we have chosen a non trivial literal from $C_2$
for resolution). 
No splitting or replacement rules apply and
 ${S'} = S \cup \{C\}$. We have
$\qrep{C_1}[\u] \cup \{\qrep{C_2}[\varone,\ldots,\varr]\} \supseteq
\{\qrep{C_1}[f(x_1,\ldots,x_n)], \allowbreak
\qrep{C_2}[\varone,\ldots,\varr]\}\pmodels 
\qrep{C}[\varone,\ldots,\varr]$.
Hence $\qrep{C_1}[\u] \cup \proj(\qrep{C_2}[\varone,\allowbreak \ldots,\varr]) \pmodels 
\allowbreak \proj(\qrep{C}[\allowbreak \varone,\allowbreak \ldots,\varr])$ and
 $\qrep{C_1}\allowbreak [\allowbreak \u\allowbreak [\allowbreak \ngrr \allowbreak \cup \allowbreak \ngrr\allowbreak [\ngrr[\gone]]] \cup 
\qrep{C_2}[\ngrr \cup \ngrr[\ngrr[\gone]]]) \pmodels 
\qrep{C}[\ngrr \cup \ngrr[\ngrr[\gone]]]$. Hence 
$\inst(S) \supseteq \instone(\qrep{C_1}) \cup \instfour(\qrep{C_2}) \pmodels
\instfour(\qrep{C})$.
\begin{itemize}
\item 
Suppose $C'_1$ is non-empty or $C'_2$ has some non-trivial literal.
Then $C$ is of type C4. 
The only trivial literals in $C[\varone,\ldots,\varr]$ are those in 
$C'_2[\varone,\ldots,\varr]$ and
hence they involve predicates from $\mathbb P$. Hence $C[\varone,\ldots,\varr]$
if of type C4 and the first statement holds.
\item 
Suppose $C'_1$ is empty and $C'_2$ has only trivial literals.
Then $C[\varone,\ldots,\varr] =
B_1[\varione] \sqcup \ldots \sqcup B_k[\varik] 
\lor D_1 \lor D_2$ 
where $1\le i_1,\ldots,i_k\le r$, and each $B_i$ is an $\epsilon$-block.
The second statement holds.
\end{itemize}

\item We do resolution between a clause $C_1[\varrone] =
C'_1[\varrone] \lor D_1 \lor \pm P(s[\varrone])$ and a clause 
$C_2[\varrone] = \mp P(t[\varrone]) \lor C'_2[\varrone] \lor D_2$,
both of type C2, and both from $S$ upto renaming, and the resolvent is
$C[\varrone]=C'_1[\varrone]\sigma\lor C'_2[\varrtwo]\sigma\lor  D_1\lor D_2$ 
where $\sigma = mgu(s[\varrone],t[\varrtwo])$ (we renamed the 
second clause before resolution). 
We know that $s[\varrone], t[\varrone] \in \ngrone$, and by ordering constraints
both $s$ and $t$ are non-trivial. By Lemma~\ref{lemma:sxty:red} 
one of the following cases holds:
\begin{itemize}
\item $\varrone\sigma = \varrtwo\sigma = \varrone$. 
$C[\varrone] = C'_1[\varrone] \lor C'_2[\varrone]$.
Hence no splitting or replacement rules apply and $S' = S \cup \{C\}$.
We have $\{C_1[\varrone],C_2[\varrone]\}\pmodels C[\varrone]$. 
Hence $\insttwo(\qrep{C_1}[\varrone]) \cup \insttwo(\qrep{C_2}[\varrone]) 
\pmodels \insttwo(\qrep{C}[\varrone]) \ni \qrep C[\varrone]$.
If $C'_1$ or $C'_2$ contains some non-trivial literal 
then $C[\varrone]$ is of type C2 and the first condition holds.
If $C'_1$ and $C'_2$ contain only trivial literals then
$C$ is of type C1 and the second condition holds.

\item $\varrone\sigma,\varrtwo\sigma \in \ngrr[\g] \subseteq \ngrr[\gone]$. 
Then every literal in $C[\varrone]$ is of the
form $\pm' Q(u)$ with $u\in \ngrone[\ngrr[\gone]]$.
No splitting or replacement rules apply and $S' = S \cup \{C\}$.
$\inst(S) \supseteq \qrep{C_1}[\ngrr[\gone]] \cup \qrep{C_2}[\ngrr[\gone]] 
\pmodels \{\qrep{C}\} = \instthree(\qrep{C})$. 
If $C'_1$ or $C'_2$ is non-empty then $C$ is of type
C3 and the first statement holds.
If $C'_1$ and $C'_2$ are empty then $C$ is of type C1 and the second statement
holds.
\end{itemize}

\item We do resolution between a clause $C_1[\varrone] = C'_1[\varrone]
\lor D_1 \lor \pm P(s[\varrone])$ of type C2, and a ground
clause $\mp P(t) \lor C'_2 \lor D_2$ of type C3,
both from $S$ upto renaming, and the resolvent is
$C = C'_1[\varrone] \sigma \lor C'_2\lor  D_1 \lor D_2$ where $\sigma$ 
is a unifier of $s[\varrone]$ and $t$. 
We know that $s[\varrone]\in \ngrone$, $t\in \ngrone[\ngrr[\gone]]$,
and by ordering constraints, $s$ is non-trivial.  We have the following cases:
\begin{itemize}
\item $t\in \gone$. Then $\varrone\sigma$ is a strict subterm of $t$ hence 
$\varrone\sigma \in \g \subseteq \ngrr[\gone]$. 
\item $t\in \ngrone[\ngrr[\gone]] \setminus \gone$. Hence we have $t=t_1[t']$
for some non-trivial $t_1[\varrone]\in \ngrone$ and some $t'\in \ngrr[\gone]$. 
Let $s' = \varrone\sigma$.
As $s[s'] = t_1[t']$ hence $s[\varrone]$ and $t_1[\varrtwo]$ have a unifier
$\sigma = \{\varrone\mapsto s', \varrtwo\mapsto t'\}$. From
Lemma~\ref{lemma:sxty:red},  one of
the following is true:
\begin{itemize}
\item $s[\varrone] = t_1[\varrone]$. Hence we
have $\varrone\sigma = s'  = t' \in \ngrr[\gone]$.
\item $\varrone\sigma_1, \varrtwo\sigma_1 \in \ngrr[\g] 
\subseteq \ngrr[\gone]$. Hence $s'\in \ngrr[\gone]$. 
\end{itemize}
\end{itemize}
In each case we have $\varrone\sigma = s'\in \ngrr[\gone]$. 
Hence all literals in $C'_1[\varrone]\sigma$ are of the
form $\pm Q(t)$ with $t\in \ngrone[\ngrr[\gone]]$. All literals
in $C'_2$ are of the form $\pm' Q(t)$ with $t\in \ngrone[\ngrr[\gone]]$.
Hence no splitting or replacement rules
apply and ${S'} = S \cup \{C\}$. 
$\inst(S) \supseteq \insttwo(\qrep{C_1}[\varrone]) \cup \instthree(\qrep{C_2}) 
\supseteq \qrep{C_1}[\ngrr[\gone]] \cup \{\qrep{C_2}\}\pmodels \{\qrep{C}\} = 
\instthree(\qrep{C})$.
If $C'_1$ or $C'_2$ is non-empty then $C$ is of type C3
and the first statement holds.
 If $C'_1$ and $C'_2$ are empty then $C$ is of
type C1 and the second statement holds.

\item
We do resolution between a clause $C_1[\varrone] = C'_1[\varrone] 
\lor D_1 \lor \pm P(s[\varrone])$ of type C2,
and a clause $C_2[\varone,\ldots,\varr] = \mp 
P\allowbreak (\allowbreak f\allowbreak (\allowbreak x_1\allowbreak ,\allowbreak \ldots\allowbreak ,\allowbreak x_n\allowbreak )) \lor C'_2[\varone,\ldots,\varr] \lor D_2$ of type C4,
both from $S$ upto renaming, and $\pm P(s[\varrone])$ and $\mp
P(\allowbreak f(\allowbreak x_1,\allowbreak \ldots,\allowbreak x_n))$ are the literals resolved upon from the respective
clauses. (By ordering constraints we have chosen a non-trivial literal to
resolve upon in the second clause). 
By  ordering constraints $s[\varrone]\in\ngrone$ is non-trivial. 
Hence we have the following two cases for $s[\varrone] = 
f(s_1[\varrone],\ldots,s_n[\varrone])$.

\begin{itemize}
\item We have some $1\le i,j\le n$ such that
$x_i=x_j$ but $s_i[\varrone]\neq s_j[\varrone]$. By Lemma~\ref{lemma:sxtx},
 the only possible unifier of the terms
$s[\varrone]$ and $f(x_1,\ldots,x_n)$ is $\sigma$ such that 
$\varrone\sigma = g$ is a ground subterm
of $s_i$ or $s_j$ and $x_k\sigma = s_k[g]$ for $1\le k\le n$.
As $s[\varrone]\in \ngrone$, we have $g \in \g$ and
each $s_k[\varrone] \in \ngrr \cup \g$. Hence $\varrone\sigma \in \g$ 
and each $x_k\sigma \in \ngrr[\g] \cup \g \subseteq \ngrr[\gone]$. 
The resolvent $C = C'_1[\varrone]\sigma \cup C'_2[\varone,\ldots,\varr]\sigma
\lor  D_1 \lor D_2$ 
is ground. 
Each literal in $C'_1[\varrone]\sigma$ is of the
form $\pm' Q(t)$ with $t\in \ngrone[\g] \subseteq \ngrone[\ngrr[\gone]]$. Each
literal in $C'_2[\varone,\ldots,\varr]\sigma$  
is of the form $\pm' Q(t)$ where the following cases can arise:
\begin{itemize}
\item $t = f'(x_{i_1},\ldots,x_{i_m})\sigma$ such that 
$\{x_{i_1},\ldots,x_{i_m}\} = \{x_1,\ldots,x_n\}$. Then 
$t = f'(s_{i_1},\ldots,s_{i_m})[g]\in \ngrone[\gone]\subseteq 
\ngrone[\ngrr[\gone]]$.
\item $t = x_k \sigma
\in \ngrr[\gone]\subseteq \ngrone[\ngrr[\gone]]$ 
for some $1\le k\le n$, where the literal
$\pm' Q(x_k)$ is from $C_2$. 
\end{itemize}
We conclude that all non-splitting 
literals in $C$ are of the form $\pm' Q(t)$ with
$t\in \ngrone[\ngrr[\gone]]$, and no splitting or replacement rules apply. 
We have ${S'} = S \cup \{C\}$.
$\inst(S) \supseteq
\insttwo(\qrep{C_1}[\varrone]) \cup \instfour(\qrep{C_2}[\varone,\ldots,\varr])
\supseteq \qrep{C_1}[\ngrr[\gone]] \cup \qrep{C_2}[\ngrr[\gone]] \pmodels \{C\}
=\instthree(C)$. 
If $C'_1$ or $C'_2$  is non-empty then $C$ is of type C3, and the first
statement holds.
If $C'_1$ and $C'_2$ are empty then $C$ of type C1 and the second
condition holds. 

\item For all $1\le i,j\le n$, if $x_i=x_j$ then
$s_i[\varrone]=s_j[\varrone]$. 
Then $s[\varrone]$ and $f(x_1,\ldots,x_n)$ have mgu $\sigma$ 
such that $x_k\sigma = s_k[\varrone]\in \ngrr\cup \g$ for $1\le k\le n$ and 
$x\sigma = x$ for $x \notin \{x_1,\ldots,x_n\}$.
The resolvent $C[\varrone] = C'_1[\varrone] \lor C'_2\sigma\lor  D_1 \lor D_2$ is a 
one-variable clause. 
$\{C_1[\varrone]\}\cup C_2[\ngrr\cup\g] \pmodels C[\varrone]$. 
All literals in $C'_1[\varrone]$ are of the form
$\pm' Q(t)$ with $t\in \ngrone$, and no replacement rules
apply on them. All literals in $C'_2[\varone,\ldots,\varr]\sigma$ 
are of the form $\pm' Q(t[\varrone])$ where the following cases can arise:
\begin{itemize}
\item $t[\varrone] = f'(x_{i_1},\ldots,x_{i_m})\sigma$ such that 
$\{x_{i_1},\ldots,x_{i_m}\} = \{x_1,\ldots,x_n\}$. Then $t[\varrone]\in 
\ngrone$. No replacement rules apply on such a literal.
\item $t[\varrone] = x_k \sigma = s_k[\varrone]\in \ngrr$ for some 
$1\le k\le n$, where the literal
$\pm' Q(x_k)$ is from $C_2$. Hence we must have $Q\in\mathbb P$. 
Let $s_k[\varrone] = t_1[\ldots[t_p[\varrone]]\ldots]$ for some $p\ge 0$ 
where each $t_i[\varrone]\in\ngr$ is non-trivial and reduced. Such a literal
is replaced by the literal $\pm' \allowbreak Qt_1\ldots t_{p-1}\allowbreak (\allowbreak t_p\allowbreak [\allowbreak \varrone\allowbreak ]\allowbreak )$ and we 
know that $t_p \in \ngr\subseteq \ngrone$.
This new clause is obtained by propositional resolution between the former
clause and clauses from $cl(\mathcal R)[\ngrr]$.
\item $t[\varrone] = x_k \sigma = s_k\in \g$ for some 
$1\le k\le n$, where the literal
$\pm' Q(x_k)$ is from $C_2$. Hence we must have $Q\in\mathbb P$. 
No replacement rules apply on such a literal.
If $C$ contains only ground literals then
this literal is left unchanged. Otherwise we perform
$\grpreds$-splitting and this literal is replaced by the
literal $- \ssp{\pm' Q(s_k)}$ and also a new clause $C'' = \ssp{\pm' Q(s_k)}
\lor \pm' Q(s_k)$ of type C3 is added to $S$. If $C'$ is the new clause
obtained by this splitting then $\qrep{C'}$ is clearly propositionally
equivalent to the former clause. Also $\qrep{C''} = \mp' Q(s_k) \lor \pm'
Q(s_k)$ is a propositionally valid statement.
\end{itemize}
We conclude that after zero or more replacement and splitting
rules, we obtain a clause $C'[\varrone]$, 
together with a set $S''$ of clauses of type C3, 
$\{\qrep C[\varrone]\} \cup cl(\mathcal R)[\ngrr]\pmodels
\{\qrep{C}[\varrone]\}$, $\emptyset \pmodels \qrep{S''}$, and
${S'} = S \cup \{C'\}\cup S''$.
$\{\qrep{C_1}[\varrone]\} \cup \qrep{C_2}[\ngrr \cup \g] 
\cup cl(\mathcal R)[\ngrr]\pmodels \qrep{C'}[\varrone]$. 
Hence $\inst(S) \supseteq \insttwo(\qrep{C_1}) \cup \instfour(\qrep{C_2})
\supseteq \{\qrep{C_1}[\varrone]\} \cup \qrep{C_1}[\ngrr[\gone]]
\cup \qrep{C_2}[\ngrr\cup\ngrr[\ngrr[\gone]]] 
\cup cl(\mathcal R)[\ngrr] \cup cl(\mathcal R)[\ngrr[\ngrr[\gone]]]
\pmodels \insttwo(\qrep{C'}) \cup \instthree(\qrep{S''}) = 
\qrep{C'}[\varrone] \cup \qrep{C'}[\ngrr[\gone]] \cup \qrep{S''}$.
If $C'$ is of type C2 or C3 then the first statement holds.
Otherwise $C'$ is  of type C1 and the second statement holds.
\end{itemize}

\item We do resolution between a clause $C_1 = C'_1 \lor D_1
\lor \pm P(s)$ and a clause $C_2 = \mp P(s) \lor C'_2 \lor D_2$, 
both ground clauses of type C3 from $S$, 
and the resolvent is $C = C'_1 \lor C'_2\lor  D_1 \lor D_2$. 
No replacement or splitting rules
apply and we have ${S'} = S \cup \{C\}$. 
$\inst(S) \supseteq \{\qrep{C_1},\qrep{C_2}\} \pmodels \instthree(\qrep C)
= \{\qrep{C}\}$.  If $C'_1$ or $C'_2$ is non-empty then $C$ is of type C3,
 and the first statement holds.
If $C'_1$ and $C'_2$ are empty then $C$ is of type C1 and the second
statement holds.

\item
We do resolution between a ground clause $C_1 = C'_1 \lor D_1 \lor \pm P(s)$ of
type C3, and a clause $C_2[\varone,\ldots,\varr] = \mp P(f(x_1,\ldots,x_n)) 
\lor C'_2[\varone,\ldots,\varr] \lor D_2$ of type C4,
both from $S$ upto renaming, and $\pm P(s)$ and $\mp
P(f(x_1,\ldots,x_n))$ are the literals resolved upon from the respective
clauses. We know that $s\in  \ngrone[\ngrr[\gone]]$. 
Hence we have the following two cases for $s$.
\begin{itemize}
\item $s \in \ngrone[\ngrr[\gone]] \setminus \gone$. 
Hence $s$ must be of the form
$f(s_1,\ldots,s_n)[g]$ for some $f(s_1,\ldots,s_n)\in \ngrone$ and some 
$g \in \ngrr[\gone]$ (The symbol $f$ is same as in the literal 
$\mp P(f(x_1,\ldots,x_n))$ otherwise this resolution step would not be 
possible). We have each $s_i \in \ngrr \cup \g$. The mgu $\sigma$ of $s$ and 
$f(x_1,\ldots,x_n)$ is such that $x_i\sigma = s_i[g]\in \ngrr[\ngrr[\gone]]$.
The resolvent $C = C'_1 \lor C'_2[\varone,\ldots,\varr]\sigma\lor  D_1 \lor D_2$
is a ground clause. All literals in $C'_1$ are of the
form $\pm' Q(t)$ with $t\in \ngrone[\ngrr[\gone]]$ hence no replacement
rules apply on them.  The literals in $C'_2[\varone,\ldots,\varr]\sigma$ are of the form $\pm' Q(t)$
where the following cases are possible:
\begin{itemize}
\item $t = f'(x_{i_1},\ldots,x_{i_m})\sigma$ where $\{x_{i_1},\ldots,x_{i_m}\}
= \{x_1,\ldots,x_n\}$. Then $f'\allowbreak (\allowbreak s_{i_1}\allowbreak ,\allowbreak \ldots\allowbreak ,\allowbreak s_{i_m}\allowbreak ) \in \ngrone$. Hence
$t \in \ngrone[\ngrr[\gone]]$. No replacement rules apply on such a literal.
\item $t = x_i\sigma \in \ngrr[\ngrr[\gone]]$ for some
$1\le i\le n$. If $t\in \ngrone[\ngrr[\gone]]$ then no replacement rules  apply
on this literal. Otherwise suppose $t\in \ngrr[\ngrr[\gone]]\setminus
\ngrone[\ngrr[\gone]]$. We have $t=t_1[\ldots[t_p[t']]\ldots]$ for some
reduced non-trivial non-ground terms $t_1,\ldots,t_p\in \ngr$ with $p\ge 0$
such that $t_1[\ldots[t_p[y]]] \in \ngrr$ and $t'\in \ngrr[\gone]$, and the
replacement strategy replaces this literal by the literal 
$\pm' Qt_1\ldots t_{p-1}(t_p[t'])$, and we know that
$t_p\in \ngr \subseteq \ngrone$ so that $t_p[t']\in \ngrone[\ngrr[\gone]]$. 
This new clause can be
obtained by propositional resolution between the former clause and  clauses
from $cl(\mathcal R)[\ngrr[\ngrr[\gone]]]$ 
\end{itemize}
We conclude that after zero or more replacement rules, we obtain a ground
clause $C'$, all of whose non-splitting 
literals are of the form $\pm' Q(t)$ with
$t\in \ngrone[\ngrr[\gone]]$, and which is obtained by propositional resolution 
from $\{C\} \cup cl(\mathcal R)[\ngrr[\ngrr[\gone]]]$. No splitting rules
apply and ${S'} = S \cup \{C'\}$.  
$\{\qrep{C_1}\} \cup \qrep{C_2}[\ngrr[\ngrr[\gone]]] \pmodels \qrep{C}$ hence
$\inst(S) \supseteq \instthree(\qrep{C_1})\cup\instfour(\qrep{C_2}) \cup
cl(\mathcal R)[\ngrr[\ngrr[\gone]]] \pmodels \instthree(\qrep{C'}) 
= \{\qrep{C'}\}$.  If $C'_1$ or $C'_2$ is non-empty then
$C$ is of type C3, and the first statement holds.
If $C'_1$ and $C'_2$ are empty then $C$ is 
of type C1 and the second statement holds.

\item $s \in \gone$. For the resolution step to be possible we must have
$s = f(s_1,\ldots,s_n)$. Each $s_i \in \g$.
 The mgu $\sigma$ of $s$ and 
$f(x_1,\ldots,x_n)$ is such that each $x_i\sigma = s_i$.
The resolvent $C = C'_1 \lor C'_2[\varone,\ldots,\varr]\sigma\lor  D_1 \lor D_2$ is a ground clause.
All literals in $C'_1$ are of the
form $\pm' Q(t)$ with $t\in \ngrone[\ngrr[\gone]]$.
 The literals in $C'_2[\varone,\ldots,\varr]\sigma$ are of the form $\pm' Q(t)$
where the following cases are possible:
\begin{itemize}
\item $t = f'(x_{i_1},\ldots,x_{i_m})\sigma$ where $\{x_{i_1},\ldots,x_{i_m}\}
= \{x_1,\ldots,x_n\}$. Then $t = f'(s_{i_1},\ldots,s_{i_m}) \in \gone
\subseteq \ngrone[\ngrr[\gone]]$.
\item $t = x_i\sigma = s_i \in \g \subseteq \ngrone[\ngrr[\gone]]$ for some
$1\le i\le n$.  
\end{itemize}
Hence all non-splitting
literals in $C$ are of the form $\pm' Q(t)$ with $t\in 
\allowbreak \ngrone\allowbreak [\allowbreak \ngrr\allowbreak [\allowbreak \gone\allowbreak ]\allowbreak ]$.
No replacement rules or splitting rules apply and
${S'} = S \cup \{C\}$. 
$\{C_1\} \cup C_2[\g] \pmodels C$ hence
$\inst(S) \pmodels \instthree(\qrep C) = \{\qrep{C}\}$. 
If $C'_1$ or $C'_2$ is non-empty then
$C$ is of type C3 and the first statement holds.
If $C'_1$ and $C'_2$ are empty then $C$ is
of type C1 and the second statement holds.
\end{itemize}

\item We do resolution between two clauses $C_1[\varone,\ldots,\varr]$ and 
$C_2[\varone,\ldots,\varr]$, both of type C4,
and both from $S$ upto renaming. First we rename the second clause as 
$C_2[\varrone,\ldots,\varrr]$ by applying the renaming $\sigma_0 =
\{\varone\mapsto\varrone,\ldots,\varr\mapsto\varrr\}$.
By ordering constraints, $C_1[\varone,\ldots,\varr] 
= C'_1\allowbreak [\allowbreak \varone\allowbreak ,\allowbreak \ldots\allowbreak ,\allowbreak \varr\allowbreak ] \allowbreak \lor \allowbreak D_1
\allowbreak \lor \allowbreak P(\allowbreak f(\allowbreak x_1,\allowbreak \ldots,\allowbreak x_n\allowbreak ))$
and $C_2[\varrone,\ldots,\varrr] = - P(f(y_1,\ldots,y_n)) \lor C'_2[\varrone,\ldots,\varrr] \lor D_2$ and the resolvent is
$C[\varone,\ldots,\varr] = C'_1[\varone,\ldots,\varr]\sigma \lor C'_2[\varrone,\ldots,\varrr]\sigma
\lor D_1 \lor D_2$ 
where, by Lemma~\ref{lemma:sequnif}, $\sigma$ is such that
$\{x_1,\ldots,x_n\}\sigma \subseteq \{x_1,\ldots,x_n\}$ and 
$y_i\sigma = x_i$ for $1\le i\le n$.
$\proj(C_1) \cup \proj(C_2) \pmodels C[\varone,\ldots,\varr]$.
Hence $\inst(S) \supseteq \instfour(\qrep{C_1}[\varone,\ldots,\varr]) \cup
\instfour\allowbreak (\allowbreak \qrep{C_2}\allowbreak [\varone\allowbreak ,\allowbreak \ldots\allowbreak ,\allowbreak \varr\allowbreak ]\allowbreak ) =
\proj(\qrep{C_1}[\varone,\ldots,\varr]) \allowbreak \cup \allowbreak \qrep{C_1}[\ngrr \allowbreak \cup \allowbreak \ngrr[\ngrr[\allowbreak \gone]]]
\cup \proj(\qrep{C_2}[\varone,\ldots,\allowbreak \varr\allowbreak ]) \allowbreak \cup\allowbreak  \qrep{C_2}[\allowbreak \ngrr\allowbreak  \cup \allowbreak \ngrr[\allowbreak \ngrr[\allowbreak \gone]]]
\allowbreak  \pmodels \allowbreak  \proj(\qrep{C}[\varone,
\allowbreak  \ldots,\allowbreak  \varr]) \cup \qrep{C}[\ngrr\cup\ngrr[\ngrr[\gone]]] = \instfour(\qrep{C}[\varone,\ldots,\varr])$.
\begin{itemize}
\item Suppose $C'_1$ or $C'_2$ has a non-trivial literal. Then $C$ is of type
C4, no replacement or splitting rules apply, 
$S' = S \cup \{C\}$ and the first statement holds.
\item Suppose $C'_1$ and $C'_2$ contain no non-trivial literal. Then $C[\varone,\ldots,\varr] = 
B_1[\varione] \sqcup \ldots \sqcup B_k[\varik]\lor D_1 \lor D_2$ 
with $1\le i_1,\ldots,i_k\le r$, each $B_i$ being an $\epsilon$-block. 
No splitting or replacement rules apply ($\epsilon$-splitting is forbidden
by $\phi_0$), and $S' = S \cup \{C\}$. The second statement holds.
\end{itemize}

\item We do a resolution step in which one of the premises is a clause from
$cl(\mathcal R)$. Every clause in $cl(\mathcal R)$ is of type C2.
Also trivially $\insttwo(C) \subseteq \inst(T)$.
Hence this case can be dealt with in the same way
as in the case  where one of the premises of resolution is a clause
of type C2.
\end{enumerate}

Next we consider factoring steps.
Factoring on a clause of type C1 or C3 is possible only if the two
involved literals are the same, hence this is equivalent to doing nothing.

\begin{enumerate}
\item We do factoring on a clause $C_1[\varrone] = C'_1[\varrone] 
\lor \pm P(s[\varrone])\lor \pm P(t[\varrone])$
of type C2, and from $S$ upto renaming. 
We know that $s[\varrone],t[\varrone]\in \ngrone$, and by 
ordering constraints $s$ and $t$ are non trivial. 
The clause obtained is
$C[\varrone] = C'_1[\varrone]\sigma \lor \pm P(s[\varrone])\sigma$ 
where $\sigma$ is a unifier of $s[\varrone]$ and $t[\varrone]$.
If $s[\varrone]\neq t[\varrone]$ 
then by Lemma~\ref{lemma:sxtx} 
$\varrone\sigma$ is a ground strict subterm of $s$ or $t$, 
hence $\varrone\sigma \in \g
\subseteq \ngrr[\gone]$. Each literal in $C$ is of the form $\pm' Q(t')$ where
$t'\in \ngrone[\ngrr[\gone]]$. Hence $C$ is of type C3. No splitting or
 replacement rules apply and ${S'} = S \cup \{C\}$.
We have $C \in C_1[\ngrr[\gone]]$.
$\inst(S) \supseteq \insttwo(\qrep{C_1}[\varrone])
\supseteq\qrep{C_1}[\varrone][\ngrr[\gone]] \supseteq\instthree(C)=\{\qrep C\}$.
The first statement holds.

\item We do factoring on a clause $C_1[\varone,\ldots,\varr]$
of type C4, and from $S$ upto renaming, 
to obtain the clause $C[\varone,\ldots,\varr]$.
By ordering constraints non-trivial literals must be chosen for factoring.
Then $C[\varone,\ldots,\varr]$ is again of type C4 and $C[\varone,\ldots,\varr]
\in \proj(C_1)$.
$\inst(S) \supseteq \instfour(\qrep{C_1}) = 
\proj(\qrep{C_1}) \cup \qrep{C_1}[\ngrr\cup
\ngrr[\ngrr[\gone]]] \pmodels \instfour(\qrep{C})$. 
The first statement holds.
\qed
\end{enumerate}
\endproof 
\restorelemmacounter

The alternative resolution procedure for testing unsatisfiability by using
succinct representations of tableaux is now defined by the rule:
$\mathcal T \mid S \absres \mathcal T \mid S \cup \{B_1 \sqcup D\} 
\mid S\cup\{B_2\}\mid\ldots \mid S \cup \{B_k\}$ 
whenever $\inst(S) \pmodels B_1\sqcup \ldots
\sqcup B_k \sqcup \qrep{D}$, each $B_i$ is an $\epsilon$-block, 
$1\le i_1,\ldots,i_k \le r$ and $D \subseteq \pm\sspreds$.
The simulation property now states:

\newcounter{lemnhsimul}
\namelem{lemnhsimul}
\def\lemnhsimulstatement{
If $S \abss T$ and $S\subordselfssplreplres{\mathcal T}$ then 
$T \absres^* {\mathcal T}'$ for some ${\mathcal T}'$ such that $\mathcal T \abss
\mathcal T'$.
}
\begin{lemma}
\label{lemma:nhsimul}
\lemnhsimulstatement
\end{lemma}
\proof
As $S\subordselfssplreplres{\mathcal T}$, we have some $S'$ such that
$S\subordselnosplreplres S'$ and $\mathcal T$ is obtained from $S'$ by
$\epsilon$-splitting steps.  From Lemma~\ref{lemma:nhresol},
one of the following cases holds.
\begin{itemize}
\item   $S' \abss S$. Then $S'$ contains only clauses of type
C1-C4 and no $\epsilon$-splitting is applicable. Hence $\mathcal T = S'
\abss S$. As $\mathcal T \abss S$ and $S \abss T$ hence
$\mathcal T \abss T$ because of transitivity of $\abss$. 
Thus $T$ is the required $\mathcal T'$.
\item  $S' = S \cup \{C\}\cup S''$, $C$ is a renaming of
$B_1[\varione] \sqcup \ldots\sqcup B_k[\varik] \sqcup D$ where each
$B_i$ is an $\epsilon$-block, $1\le i_1,\ldots,i_k\le r$, $D \subseteq\pm
\sspreds$, $\inst(S) \pmodels \qrep C$ and $S''$ is a set of clauses of type
C3 and $\emptyset \pmodels \qrep{S''}$. We have $\mathcal T =
S \cup S'' \cup \{B_1 \sqcup D\} 
\mid S\cup S'' \cup \{B_2\}\mid\ldots\mid S \cup S'' \cup \{B_k\}$.
We have $S \cup S'' \cup \{B_1\sqcup D\} \abss T \cup \{B_1 \sqcup D\}$ and
$S \cup S'' \cup \{B_i\} \abss T \cup \{B_i\}$ for $1\le i\le k$.
We show that the required $\mathcal T'$ is $T \cup \{B_1\sqcup D\} \mid
T \cup \{B_1\} \mid \ldots \mid S \cup S'' \cup \{B_k\}$.
As $S \abss T$ hence $\inst(T) \pmodels \inst(S)\pmodels \qrep C$. 
Hence $T \absres \mathcal T'$.\qed
\end{itemize}
\endproof

Hence as for flat clauses we obtain:

\newcounter{thcsat}
\nameth{thcsat}
\def\thcsatstatement{
Satisfiability for  the class $\c$ is NEXPTIME-complete.
}
\savethcounter
\putth{thcsat}
\begin{theorem}
\label{theorem:csat}
\thcsatstatement
\end{theorem}
\proof 
Let $S$ be a finite set in $\c$ whose satisfiability we want to show.
We proceed as in the proof of Theorem~\ref{theorem:flatsat}. Wlog if $C\in S$
then $C$ is either a complex clause or
a one-variable clause. Clearly
$S$ is satisfiable iff $S \cup cl(\mathcal R)$ is satisfiable. At the beginning
we apply 
the replacement steps using $\mathcal R$ as long as  possible and then
$\grpreds$-splitting as long as possible.
Hence wlog all clauses in $S$ are of type C1-C4.
Then we non-deterministically add a certain number of clauses
of type C1 to S. Then we check that the resulting set  $S'$
does not contain $\empcl$, and is saturated in the sense
that: if $C = B_1[\varione]\sqcup \ldots \sqcup
B_k[\varik]\sqcup D$, each $B_i$
is an $\epsilon$-block, $1\le i_1,\ldots,i_k\le r$, $D\subseteq\pm\grpreds$,
 and $B_j[\varrone]
\notin S'$ for $1\le j\le k$, then $\inst(S') \pnmodels \qrep{C}$.
There are exponentially many such $C$ to check for since the number of
splitting literals in polynomially many. The size of
$\inst(S')$ is exponential.\qed
\endproof

\section{The Horn Case}
\label{sec:horn}

We show that in the Horn case, the upper bound can be improved to DEXPTIME.
The essential idea is that propositional satisfiability of Horn clauses is
in PTIME instead of  NPTIME. But now we need to eliminate the 
use of tableaux altogether. 
To this end, we replace the $\epsilon$-splitting rule of 
Section~\ref{sec:nh} by splitting-with-naming. Accordingly we instantiate the
set $\sspreds$ used in Section~\ref{sec:nh} as 
$\sspreds = \grpreds \cup\epspreds$ where
$\epspreds = \{\ssp{C} \mid C \textrm{ is a non-empty negative 
$\epsilon$-block with predicates from }\mathbb P\}$. 
We know that binary resolution and factorization on Horn clauses produces 
Horn clauses.  Replacements on Horn clauses
using the rules from $\mathcal R$ produces Horn clauses.
$\epspreds$-splitting on Horn clauses produces Horn clauses. E.g. clause
$P(\varone) \lor -Q(\varone) \lor -R(\vartwo)$ produces
$P(\varone) \lor -Q(\varone) \lor -\ssp{-R(\vartwo)}$ and $\ssp{-R(\vartwo)}
\lor - R(\vartwo)$.
$\grpreds$-splitting on $P(f(x)) \lor -Q(a)$ produces
$P(f(\varone)) \lor -\ssp{-Q(a)}$ and $\ssp{-Q(a)} \lor -Q(a)$ which are Horn.
However $\grpreds$-splitting on $C = -P(f(\varone)) \lor Q(a)$ produces
$C_1 = -P(f(\varone)) \lor -\ssp{Q(a)}$ and $C_2 = \ssp{Q(a)} \lor Q(a)$. $C_2$
is not Horn. However $\qrep{C_1} = C$ and $\qrep{C_2} = -Q(a)\lor Q(a)$ are
Horn. Finally, as $\epspreds$ has exponentially many atoms, 
we must restrict their occurrences in clauses. Accordingly, for $1\le i\le4$,
define clauses of type C$i$' to be clauses $C$ of the type C$i$, 
such that $\qrep{C}$ is Horn and has at most $r$ negative literals 
from $\epspreds$.
($\qrep{C}$ is defined as before, hence it leaves atoms from $\epspreds$ 
unchanged).  Now the $\sspreds$-splitting-replacement strategy $\phi_h$ 
first applies
the replacement steps of Section~\ref{sec:nh} as long as possible, then applies
$\grpreds$-splitting as long as possible and then applies 
$\epspreds$-splitting as long as possible.
Succinct representations are now defined as: $S \absh T$ iff for each
$C\in S$, $C$ is of type C$i$' and satisfies P$i_T$ for some $1\le i\le 4$.
The abstract resolution procedure is defined as:
$T \absresh T \cup \{B_1\lor -q_2\lor \ldots \lor -q_k 
\sqcup D\sqcup E\} \cup \{\ssp{B_i} \lor B_i \mid 2\le i\le k\}$ whenever
$\inst(T) \pmodels \qrep{C}$,
$C = B_1[\varione] \sqcup \ldots \sqcup B_k[\varik]\sqcup D\sqcup E$,
$\qrep{C}$ is Horn, $1\le i_1,\ldots,i_k\le r$,
$B_1$ is an $\epsilon$-block, $B_i$ is a negative $\epsilon$-block and
$2\le i\le k$, $D \subseteq \pm \grpreds$
and $E\subseteq \pm\epspreds$ such that if $k=1$ then $E$ has at most 
$r$ negative literals, and if $k>1$ then $E$ has no negative literal.
The $\abss$ and $\absres$ relations are as in Section~\ref{sec:nh}.

\newcounter{lemhresol}
\namelem{lemhresol}
\def\lemhresolstatement{
If $S \absh T$ and $S \subordselfssplreplresh S_1$ then
$T \absresh^* T_1$ and $S_1\absh T_1$ for some $T_1$.
}
\begin{lemma}
\label{lemma:hresol}
\lemhresolstatement
\end{lemma}
\proof
Let $\phi_0$ be as in Section~\ref{sec:nh}.
As $S \subordselfssplreplresh S_1$ hence we have some $S'$ such that 
$S \subordselnosplreplres S'$ and
$S_1$ is obtained from $S'$ by applying $\epspreds$-splitting steps.
As discussed above, all clauses $C\in S_1\cup S'$ 
 are such that $\qrep{C}$ is also Horn.
If $S'$ is obtained by resolving upon splitting literals, 
then one of the premises must
be just a positive splitting literal. The other premise has at most
$r$ literals of the form $-q$ with $q\in\epspreds$, 
hence the resolvent has at most $r$  literals of the form $-q$ with
$q\in\epspreds$. In case non-splitting literals are resolved upon then the
premises cannot have any negative splitting literal and the resolvent has
no negative splitting literal.  $\grpreds$-splitting does not create
literals from $\pm\epspreds$. Hence all clauses in $S'$ have at most $r$
literals of the form $-q$ with $q\in\epspreds$.
Now by Lemma~\ref{lemma:nhresol}, one of the following conditions holds.
\begin{itemize}
\item $S'\abss S$. Then $\epspreds$-splitting is not applicable on clauses in
$S'$ and $S_1 = S' \abs S$. From transitivity of $\abs$ we have
$S_1 \abs T$.  Then from the above discussion we conclude that $S_1 \absh T$.

\item $S' = S \cup \{C\} \cup S''$, $C$ is a renaming of
$B_1[\varione]\sqcup \ldots \sqcup
B_k[\varik]\sqcup D$, each $B_i$ is an $\epsilon$-block, 
$1\le i_1,\ldots,i_k\le r$, $D \subseteq\pm\sspreds$,
$\inst(S) \pmodels \qrep C$, and $S''$ is a set of clauses of type C3 and
$\emptyset \pmodels \qrep{S''}$. Also if $k\ge 2$ then $D$ has no 
literals $-q$ with  $q\in \epspreds$.  As $C$ is Horn,
wlog $B_i$ is negative for $i\ge 2$. Hence
$S_1 = S' \cup \{B_1 \lor -q_2\lor \ldots \lor -q_k\sqcup D\} \cup 
\{\ssp{B_i} \cup B_i \mid 2\le i\le k\}$.
We show that the required $T_1$ is $T \cup 
\{B_1 \lor -q_2\lor \ldots \lor -q_k\sqcup D\} \cup \{\ssp{B_i} \cup B_i \mid
2\le i\le k\}$. Each $\ssp{B_i} \cup B_i$ is of type C1'. 
As $C \in S'$ hence $D$ has at most $r$ literals $-q$ with
$q\in\epspreds$.  Hence if $k=1$ then
$B_1 \lor -q_2\lor \ldots \lor -q_k\sqcup D$ is also of type C1'.
If $k\ge 2$ then $D$ has no negative literals $-q$ with $q\in\epspreds$,
and $B_1 \lor -q_2\lor \ldots \lor -q_k\sqcup D$ is again of type C1' since
$k\le r$.  As $S \absh T$ we have $\inst(T) \pmodels \inst(S) \pmodels \qrep C$.
Hence $T \absresh T_1$. Finally, clearly $S_1 \abss T_1$ hence
$S_1 \abss_h T_1$.\qed
\end{itemize}
\endproof

Now for deciding satisfiability of a set of flat and one-variable clauses we 
proceed as in the non-Horn case. But now instead of 
non-deterministically adding clauses,
we compute a sequence $S = S_0 \absresh S_1 \absresh S_2 \ldots$ starting from 
the given set $S$, and proceeding don't care non-deterministically, till no more clauses can be 
added, and then check whether
$\empcl$ has been generated.
The length of this sequence is at most exponential.
Computing $S_{i+1}$ from $S_i$ requires at most exponential time because
the number of possibilities for $C$ in the definition of $\absres$ above
is exponential.
(Note that this idea of 
$\epspreds$-splitting would not have helped in the non-Horn 
case because we cannot
bound the number of positive splitting literals in a clause in the
non-Horn case, whereas Horn clauses by definition have at most one positive 
literal).  Also note  that APDS can be encoded using flat Horn clauses. Hence:

\begin{theorem}
\label{theorem:chsat}
Satisfiability for the classes $\ch$ and $\flath$ is DEXPTIME-complete.
\end{theorem}

Together  with Theorem~\ref{th:hardness}, this gives us optimal complexity for
protocol verification:
\begin{theorem}
Secrecy of cryptographic protocols with single blind copying, with bounded
number of nonces but unbounded number of sessions is DEXPTIME-complete.
\end{theorem}

\subsection{Alternative Normalization Procedure} 

While 
Theorem~\ref{theorem:chsat} gives us the optimum complexity for the Horn case,
we outline here an alternative normalization procedure for deciding 
satisfiability in the Horn case, in the style of~\cite{seidl:h1}. 
Our goal is to show that the Horn case can be dealt with using simpler 
techniques. This may also be interesting for implementations, since it avoids 
exhaustive generation of instantiations of clauses. 
Since we already have the optimum complexity from 
Theorem~\ref{theorem:chsat}, we restrict ourselves to giving only the
important ideas here.  Define 
{\em normal} clauses to be clauses which have no function symbol in the
body, have no repetition of variables in the body, and have no variables in the
body other than those in the head. Sets of normal definite
clauses involving unary predicates can be thought of
as generalizations of tree automata, by adopting the convention that term $t$ is
{\em accepted} at {\em state} $P$ iff atom $P(t)$ is reachable.
I.e. states are just unary predicates. {\em (Intersection-)emptiness}
and {\em membership} properties are defined as usual.

\begin{lemma}
Emptiness and membership properties are decidable in polynomial time for 
sets of normal definite clauses.
\end{lemma}
\proof
Let $S$ be the set of clauses.
To test emptiness of a state $P$, we remove arguments of predicate symbols
in clauses, and treat predicates as proposition symbols. Then we add the clause
$- P$ and check satisfiability of the resulting propositional Horn clause set.

To test if $t$ is accepted at $P$, 
let $T$ be the set of subterms of $t$.
Define a set $S'$ of clauses as follows. If $Q(s)\lor -Q_1(x_1)\lor \ldots \lor
-Q_n(x_n) \in S$ and $s\sigma \in T$ for some substitution $\sigma$ then
we add the Horn clause $Q(s\sigma) \lor -Q_1(x_1\sigma) \lor \ldots \lor
-Q_n(x_n\sigma)$   to $S'$. Finally we add $- P(t)$ to $S'$
and test its unsatisfiability.  $S'$ is computable in polynomial time. 
Also $S'$ has only ground clauses, hence satisfiability is equivalent to
propositional unsatisfiability, by treating each ground literal as a 
propositional symbol.\\
\strut \hfill  \qed
\endproof

The intuition behind the normalization procedure is as follows. 
We use new states which are sets $\{P_1,P_2,\ldots,\}$, where $P_1,P_2,\ldots$
are states in the given clauses set.
The state $\{P_1,P_2,\ldots,\}$ represents intersection of the states
$P_1,P_2,\ldots$. These new states are denoted by $p,q,p_1,\ldots$.
The states $P$ in clauses are replaced by $\{P\}$.  We try to make
non-normal clauses redundant by resolving them with normal clauses.
Hence a clause $C \lor -p(t)$, where $t$ has some function symbol, is resolved
with a normal clause $p(s) \lor D$ to obtain a clause $C\sigma \lor D\sigma$ 
where $\sigma=mgu(s,t)$. Normal clauses $p(s) \lor C$ and $p(t) \lor D$ are
used to produce clause $(p\cup q)(s\sigma) \lor C\sigma \lor D\sigma$ where
$\sigma = mgu (s,t)$. In this process if we get a clause $C \lor -p(t)$
where $t$ is ground, then either $t$ is accepted at $p$ using the normal 
clauses and we remove the literal $-p(t)$ from the clause, or $t$ is not 
accepted at $p$ using the normal clauses, and we reject the clause.
From clauses $C \lor -p(x) \lor -q(x)$ we derive the clause 
$C \lor -(p\lor q)(x)$. If a clause $p(x_1) \lor -q(x_1) \lor -q_1(x_2)\lor 
\ldots\lor -q_n(x_n)$ is produced where the $x_i$ are mutually distinct, then
either each $q_i$ is non-empty using the normal clauses and we replace this
clause by $p(x) \lor -q(x)$, or we reject this clause.  The normal clauses
$p(x) \lor -q(x)$ and $q(t) \lor C$ produce the clause $q(t) \lor C$.
Replacement rules are also applied as in the non-Horn case. We continue this
till no more new clauses can be produced. Then we remove all non-normal
clauses. We claim that this process takes exponential time and 
each state $p$ in the resulting clause set accepts exactly the terms accepted
by each $P\in p$ in the original clause set. This also gives us a DEXPTIME
algorithm for the satisfiability problem for the class $\c$.

\begin{example}
Consider the set $S = \{C_1,\ldots,C_5\}$ of clauses where
$$\begin{array}{l r l}
C_1 = & P (a) & \\
C_2 = & Q (a) & \\
C_3 = & P(f(g(\varone,a),g(a,\varone),a)) & \lor -P (\varone)\\
C_4 = & P(f(g(\varone,a),g(a,\varone),b)) & \lor -P (\varone)\\
C_5 = & R (\varone) & \lor -P (f(\varone,\varone,\vartwo)) \lor -Q(\vartwo)
\end{array}$$
We first get the following normal clauses.
$$\begin{array}{l r l}
C'_1 = & \{P\} (a) & \\
C'_2 = & \{Q\} (a) & \\
C'_3 = & \{P\}(f(g(\varone,a),g(a,\varone),a)) & \lor -\{P\} (\varone)\\
C'_4 = & \{P\}(f(g(\varone,b),g(a,\varone),b)) & \lor -\{P\} (\varone)\\
\end{array}$$
The clause
$$C'_5=\{R\}(\varone)  \lor -\{P\} (f(\varone,\varone,\vartwo)) \lor -\{Q\}(\vartwo)$$
is not normal. Resolving it with $C'_3$ gives the clause
$$\{R\}(g(a,a)) \lor -\{P\}(a) \lor -\{Q\}(a)$$ 
As $a$ is accepted at $\{P\}$ and $\{Q\}$ using the
normal clauses $C'_1$ and $C'_2$, hence we get a new normal clause 
$$C_6 = \{R\}(g(a,a))$$
Resolving $C'_5$ with $C'_4$ gives 
$$\{R\}(g(a,a)) \lor -\{P\}(a) \lor -\{Q\}(b)$$
But $b$ is not accepted at $\{Q\}$ using the normal clauses hence this clause
is rejected. Finally $C'_1$ and $C'_2$ also give the normal clause
$$C_7 = \{P,Q\}(a)$$
The resulting set of normal clauses is $\{C'_1,\ldots,C'_4,C_6,C_7\}$.

\end{example}

\section{Conclusion}

We have proved DEXPTIME-hardness of  secrecy for cryptographic 
protocols with single blind copying,
and have improved the upper bound from 3-DEXPTIME to DEXPTIME.  
We have improved the 3-DEXPTIME upper bound for satisfiability for the class 
$\c$ to
NEXPTIME in the general case and DEXPTIME in the Horn case, which match
known lower bounds. For this we have
invented new resolution techniques like ordered resolution with 
splitting modulo propositional reasoning, ordered literal replacements and
decompositions of one-variable terms. As byproducts we obtained optimum
complexity for several fragments of $\c$ involving flat and one-variable
clauses. Security for several other decidable classes of protocols 
with unbounded number of sessions and bounded number of nonces
is in DEXPTIME, suggesting that 
DEXPTIME is a reasonable 
complexity class for such  classes of protocols.

\bibliographystyle{abbrv}
\bibliography{fov}

\appendix

\section{Proofs of Section~\ref{sec:onevar}}
\label{appone}

We use the following unification algorithm, due to
Martelli and Montanari.
  It is described by the
  following rewrite rules on finite multisets of equations between terms; we
  let $M$ be any such multiset, and comma denote multiset union:
  \begin{description}
  \item[(Delete)] $M, u \doteq u \to M$
  \item[(Decomp)] $M, f (u_1, \ldots, u_n) \doteq f (v_1, \ldots, v_n) \to M,
    u_1\doteq v_1, \ldots, u_n\doteq v_n$
  \item[(Bind)] $M, x \doteq v \to M [x:=v], x \doteq v$ provided $x$ is not
    free in $v$, but is free in $M$.
  \item[(Fail1)] $M, x \doteq v \to \bot$ provided $x$ is free in $v$ and
$x\neq v$.
   \item[(Fail2)] $M, f(u_1,\ldots,u_m) \doteq g(v_1,\ldots,v_n) \to \bot$
provided $f\neq g$.
  \end{description}
  We consider that equations $u \doteq v$ are unordered pairs of terms $u, v$,
  so that in particular $u \doteq v$ and $v \doteq u$ are the same equation.
  $\bot$ represents failure of unification.
  If $s$ and $t$ are unifiable, then this rewrite process terminates, starting
  from $s\doteq t$, on a so-called solved form $z_1 \doteq u_1, \ldots, z_k
  \doteq u_k$; then $\sigma = \{z_1\mapsto u_1, \ldots, z_k\mapsto u_k\}$ 
  is an mgu of $s\doteq t$.

\begin{lemma}
\label{lemma:sxty}
Let $s[x]$ and $t[y]$ be two non-ground non-trivial
one-variable terms, and $x\neq y$.
Let $U$ be the set of non-ground strict subterms of $s$ and $t$ and
let $V$ be the set of ground strict subterms of $s$ and $t$. 
If $s[x]$ and $t[y]$ are unifiable then they have a mgu $\sigma$ such that
one of the following is true:
\begin{itemize}
\item $\sigma = \{x\mapsto u[y]\}$ where $u\in U$.
\item $\sigma = \{y\mapsto u[x]\}$ where $u\in U$.
\item $\sigma = \{x\mapsto u, y\mapsto v\}$ where $u,v\in U[V]$.
\end{itemize}
\end{lemma}
\proof 
Note that $V \subseteq U[V]$ since $U$ contains the trivial terms also.
We use the above unification algorithm.
We start with the multiset $M_0 = s\doteq t$. We claim that if $M_0 \to^+ M$
then $M$ is of one of the following forms:
\begin{enumerate}
\item $s_1[x]=t_1[y],\ldots,s_n[x]=t_n[y]$, where each $s_i,t_i\in U\cup V$,
some $s_i \in U$ and some $t_j\in U$. 
\item $s_1[u[y']]=t_1[y'],\ldots,s_n[u[y']]=t_n[y'],x'=u[y']$ where $u\in U$,
each $s_i,t_i\in U\cup V$, 
$x'\in\{x,y\}$ and $y'\in\{x,y\}\setminus\{x'\}$.
\item $s_1[u]=t_1[y'],\ldots,s_n[u]=t_n[y'],x'=u$ where $u \in V$,
each $s_i,t_i\in U\cup V$, some $t_i\in U$,
$x'\in\{x,y\}$ and $y'\in\{x,y\}\setminus\{x'\}$.
\item $M', x = u, y=v$ where $u,v\in U[V]$, and no variables occur in $M'$.
\item  $\bot$.
\end{enumerate}

As $s$ and $t$ are non-trivial, and $x$ and $y$ are distinct, hence 
(Delete) and (Bind)  don't apply on $M_0$. Applying (Decomp)
on $M_0$ leads us to type (1).
Applying (Fail1) or (Fail2) on any $M$ leads us to $\bot$.
Applying (Delete) and (Decomp)  on type (1) keeps us in type (1).
Applying (Bind) on type (1) leads to type (2) or (3) depending on whether
the concerned variable is replaced by a non-ground or ground term.
Applying (Delete) on type (2) leads to type (2) itself. Applying (Decomp)
on type (2) leads to type (2) itself. (Bind) applies on $M$ of type (2) only
if $M$ contains some $y'\doteq v$ where $v$ is ground. We must have
$v \in V$. The result is of type (4).
Applying (Delete) and (Decomp) rules on type (3) leads to type (3) itself.
(Bind) applies on $M$ of type (3) only if $M$ contains some $y'\doteq v$
where $v$ is ground. We must have $v\in U[V]$. 
The result is of type (4).
Applying (Delete) and (Decomp) on type (4) leads to type (4) itself, and
(Bind) does not apply.

Now we look at the solved forms. Solved forms of type (1) are of the
form either $x\doteq u[y]$ with $u\in U$, or $y\doteq u[x]$ with $u\in U$,
or $x\doteq u,y\doteq v$ with $u,v\in V\subseteq U[V]$.
$M$ of type (2) is in solved form only if $n=0$. Hence the solved forms
are again of the form $x\doteq u[y]$ or $y\doteq u[x]$ with $u\in U$.
$M$ of type (3) is in solved form only if $n=1$, hence $M$ is of the form
$x\doteq u,y\doteq v$ with $u,v\in U[V]$. Solved forms of type (4) are again of
type $x\doteq u,y\doteq v$ with $u,v\in U[V]$ (i.e. $M'$ is empty).
\qed
\endproof

\savelemmacounter
\putlem{lemsxtyred}
\begin{lemma}
\lemsxtyredstatement
\end{lemma}
\proof 
By Lemma~\ref{lemma:sxty}, $s[x]$ and $t[y]$ have a mgu $\sigma'$ such that
one of the following is true:
\begin{itemize}
\item $\sigma' = \{x\mapsto u[y]\}$ where $u\in U$. We have $s[u[y]] = t[y]$.
As $t$ is reduced, this is possible only if $u$ is trivial. Hence
$s[y] = t[y]$, so $s[x] = t[x]$. This is a contradiction.
\item $\sigma' = \{y\mapsto u[x]\}$ where $u\in U$. This case is similar to the
previous case.
\item $\sigma' = \{x\mapsto u, y\mapsto v\}$ where $u,v\in U[V]$.
As $\sigma'$ is the  mgu and maps $x$ and $y$ to ground terms, hence
$\sigma=\sigma'$.\qed
\end{itemize}
\endproof 
\restorelemmacounter

\savelemmacounter
\putlem{lemsxtx}
\begin{lemma}
\lemsxtxstatement
\end{lemma}
\proof 
We use the above unification algorithm. We start with the multiset $M_0
= s[x]=t[x]$. If $M_0 \to^+ M$ then $M$ is of one of the following forms: 
\begin{enumerate}
\item $s_1[x]=t_1[x],\ldots,s_n[x]=t_n[x]$ where each $s_i$ is a strict
subterm of $s$ and each $t_i$ is a strict subterm of $t$
\item $M, x = u$ where $u$ is a ground strict subterm of $s$ or $t$, and no
variables occur in $M$
\item $\bot$.
\end{enumerate}

Then it is easy to see that the only possible solved form is $x\doteq u$
where $u$ is a ground strict subterm of $s$ or $t$.\qed
\endproof 
\restorelemmacounter





\section{Proofs of Section~\ref{sec:nh}}
\label{apptwo}

\savethcounter
\putth{thcompl}
\begin{theorem}
\thcomplstatement
\end{theorem}
\proof 
A {\em standard Herbrand interpretation} is a Herbrand interpretation
$\mathcal H$ such that $\ssp{C} \in \mathcal H$ iff $\mathcal H$ does not
satisfy $C$. This leads us to the notion of {\em standard satisfiability}
as expected. The given set $S$ of $\mathbb P$-clauses is satisfiable iff
it is standard-satisfiable. Ordered resolution, 
factorization and splitting preserve satisfiability in any given Herbrand
interpretation, and $\sspreds$-splitting preserves satisfiability in any
given standard-Herbrand interpretation. Also
if $T \replres T'$ then $T \cup cl(\mathcal R)$ is satisfiable in a 
Herbrand interpretation iff
$T'\cup cl(\mathcal R)$ is satisfiable in that interpretation. 
This proves correctness: if
$S\atordselfssplreplres^* \mathcal T$ and $\mathcal T$ is closed then
$S\cup cl(\mathcal R)$ is unsatisfiable.

For completeness
we replay the proof of~\cite{JGL:SOresol} for
ordered resolution with selection specialized to our
case, and insert the arguments required for the replacement rules.
Since $\atlt$ is enumerable, hence we
have an enumeration $A'_1,A'_2,\ldots$ of all ground
atoms such that if $A'_i \atlt A'_j$ then $i < j$. 
Also there are only finitely many splitting atoms in
$\sspreds$, all of which are smaller
than non-splitting atoms. Hence
the set of all (splitting as well as non-splitting) atoms
can be enumerated as $A_1,A_2,\ldots$ such that if $A_i\atltss A_j$ then $i<j$.
Clearly all the splitting atoms occur
before the non-splitting atoms in this enumeration.
Consider the infinite binary tree $\mathbb T$ whose nodes are
literal sequences of the form 
$\pm_1 A_1 \pm_2 A_2 \ldots \pm_k A_k$ for $k\ge0$. 
The two successors of the node $N$
are $N+A_{k+1}$ (the left child) and $N-A_{k+1}$ (the right child). 
If $k=0$ then $N$ is a root node. Furthermore we write
$-N = \mp_1 A_i \mp_2 A_2 \ldots \mp_k A_k$. A clause {\em fails} at a node
$N$ if there is some ground substitution $\sigma$ such that for every literal
$L \in C$, $L \sigma$ is in $-N$.
For any set $T$ of clauses define $\mathbb T_T$ as the tree obtained from
$\mathbb T$ by deleting the subtrees below all nodes of $\mathbb T$ where
some clause of $T$ fails. 
A failure-witness for a set $T$ of clauses 
is a tuple $(\mathbb T',C_\bullet,\theta_\bullet)$ such that 
$\mathbb T' = \mathbb T_T$ is finite, $C_N$ is a clause for each leaf node $N$ 
of $\mathbb T'$, and $\theta_N$ is a ground substitution for each leaf node $N$
of $\mathbb T'$ such that for $-N$  contains every 
$L\in C_N\theta_N$. We define $\nu(\mathbb T')$ as the number of nodes
in $\mathbb T'$.  For any failure witness of the form
$(\mathbb T',C_\bullet,\theta_\bullet)$ and for any leaf node 
$N=\pm_1 A_1 \pm_2 A_2 \ldots \pm_k A_k$ of 
$\mathbb T'$, define $\mu_1 (C_N,\theta_N)$ as follows:\\
-- If $C_N \notin cl(\mathcal R)$ then $\mu_1(C_N,\theta_N)$ is
the multiset of integers which contains the integer $i$ as 
many times as there are literals 
$\pm A'\in C_N$ such that $A'\theta_N = A_i$.\\
-- If $C_N \in cl(\mathcal R)$ then $\mu_1(C_N,\theta_N)$ is the
empty multiset.\\
 We define
$\mu^-(\mathbb T',C_\bullet,\theta_\bullet)$ as the multiset of
the values $\mu_1 (C_N,\theta_N)$ where $N$ ranges over all leaf nodes of 
$\mathbb T'$. We define $\mu (\mathbb T',C_\bullet,\theta_\bullet) =
(\nu(\mathbb T'),\mu^-(\mathbb T',C_\bullet,\theta_\bullet))$. We consider the
lexicographic ordering on pairs, i.e. $(x_1,y_1)< (x_2,y_2)$ iff either
$x_1 < x_2$, or $x_1=x_2$ and $y_1< y_2$.
Since $S \cup cl(\mathcal R)$ is unsatisfiable, from K\"onig's Lemma: 
\begin{lemma}
$S\cup cl(\mathcal R)$ has a failure witness.
\end{lemma}

\begin{lemma}
If $T$ has a failure witness $(\mathbb T_T, C_\bullet,\theta_\bullet)$ 
such that $\mathbb T_T$ is not just the root node, then there is some $T'$
with a failure witness $(\mathbb T_{T'},C'_\bullet,\theta'_\bullet)$ such that
$T \atordselres T'$ and $\mu(\mathbb T_{T'},C'_\bullet,\theta'_\bullet) <
\mu(\mathbb T_T, C_\bullet,\theta_\bullet)$. 
\end{lemma}
\proof 
In the following the notion of mgu is  generalized and we write
$mgu(s_1\doteq \ldots \doteq s_n)$ for the most general substitution which
makes $s_1,\ldots,s_n$ equal.
We iteratively define a sequence $R_0,R_1,\ldots$ of nodes, none of which is a
leaf node. $R_0$ is the empty sequence which is not a leaf node.
Suppose we have already defined $R_i$. As $R_i$ is not a leaf node, $R_i$
has a descendant $N_i$ such that $N_i - B_i$ is rightmost leaf node in the
subtree of $\mathbb T_T$ rooted at $R_i$. 
\begin{itemize}
\item[(1)] If $B_i$ is a non-splitting atom then stop the iteration.
\item[(2)] Otherwise $B_i$ is a splitting atom.
\begin{itemize}
\item[(2a)]
 If the subtree rooted at $N_i + B_i$ has some leaf node $N$ such that
 $- B_i \in C_N$ then stop the iteration.
\item[(2b)] Otherwise $N_i + B_i$ cannot be a leaf node. 
Define $R_{i+1} = N_i + B_i$ and continue the iteration.
\end{itemize}
\end{itemize}

$\mathbb T_T$ is finite hence the iteration terminates. Let $k$ be the
largest integer for which $R_k$, and hence $N_k$ and $B_k$ are defined.
For $0\le i\le k-1$, $B_i$ is a splitting literal. The only positive
literals in the sequence $N_k$ are from the set $\{B_0,\ldots,B_{k-1}\}$.
$N_k-B_k$ is a leaf node of $\mathbb T_T$.


Suppose the iteration stopped in case (1) above. Then $N_k$ has some descendant
$N$ such that its two children $N-B$ and $N+B$ are leaf nodes of $\mathbb T_T$,
and $B$ is a non-splitting literal. As $B_k$ is a non-splitting literal,
no negative splitting literals are present in $C_{N-B}$ or $C_{N+B}$.
$C_{N-B}$ is of the form $C_1 \lor B'_1 \lor \ldots \lor B'_m (m\ge 1)$
such that $B'_1\theta_{N-B} = \ldots = B'_m\theta_{N-B} = B$
and each literal in $C_1\theta_{N-B}$ is present in
$-N$. The literals $B'_1,\ldots,B'_m$ are then maximal in $C_{N-B}$
and can be selected for resolution.
$C_{N+B}$ is of the form $C_2 \lor -B''_1 \lor \ldots \lor -B''_n (n\ge 1)$
such that $B''_1\theta_{N+B} = \ldots = B''_n\theta_{N+B} = B$
and each literal in $C_2\theta_{N+B}$ is present in
$-N$. The literals $B''_1,\ldots,B''_n$ are then maximal in $C_{N+B}$
and can be selected for resolution. 
We assume that $C_{N-B}$ and $C_{N+B}$ 
are renamed apart so as not to share variables.
Let $\theta$ be a ground substitution which maps each
$x\in \fv(C_{N-B})$ to $x\theta_{N-B}$ and $x\in \fv(C_{N+B})$ to $x\theta_{N+B}$.
We have $B'_1\theta = \ldots = B'_m\theta = B''_1\theta = \ldots = B''_n\theta$.
Then $\sigma = mgu (B'_1\doteq \ldots \doteq B'_m\doteq B''_1 \doteq \ldots 
\doteq B''_n)$ exists. Hence we have some ground substitution $\theta'$ such that
$\sigma\theta'=\theta$.
Hence by repeated applications of the ordered factorization and ordered binary
resolution rule, 
we obtain the resolvent
$C = C_1\sigma \lor C_2\sigma$, and $T \atordselres T' = T \cup \{C\}$.
We have $C\theta' = C_1\theta \lor
C_2\theta$. Hence $C$ fails at node $N$. 
Then $\mathbb T_{T'}$ is finite
and $\nu(\mathbb T_{T'}) < \nu(\mathbb T_{T})$. Hence by choosing 
any $C'_\bullet$ and $\theta'_\bullet$ such that $(\mathbb T_{T'},C'_\bullet,
\theta'_\bullet)$ is a failure witness for $T'$, we have
$\mu(\mathbb T_{T'},C'_\bullet,\theta'_\bullet) < 
\mu(\mathbb T_T,C_\bullet,\theta_\bullet)$.

If the iteration didn't stop in  case (1) but in case (2a)
then it means that $B_k$
is a splitting literal. Then $C_{N_k-B_k} = C_1 \lor +B_k$ (with
$B_k \notin C_1$).
$C_1$ has no negative splitting literals. Hence the only literals in $C_1$ 
are positive splitting literals. Hence the literal $B_k$ can be chosen from
$C_{N_k-B_k}$ for resolution.  The subtree rooted at $N_k + B_k$ has some
leaf node $N$ such that $-B_k \in C_N$. Then $C_N = C_2 \lor -B_k$ 
(and $-B_k\notin C_2$).  Hence
$-B_k$ can be selected from $C_N$ for resolution. We obtain the resolvent
$C_2 \lor C_1$ which fails at $N$. Let $T' = T \cup \{C_1 \lor C_1\}$.
We have $\nu(\mathbb T_{T'}) \le \nu(\mathbb T_T)$.
If $N'$ is the highest ancestor of $N$ where $C_2 \lor C_1$ fails then
$N'$ is a leaf of $\mathbb T_{T'}$ and we define $C'_{N'} = C_2 \lor C_1$
and $\theta'_{N'} = \theta_N$. We have 
$\mu_1(C'_{N'},\theta'_{N'}) < \mu_1(C_N,\theta_N)$ since all 
literals in $C_1$ are splitting literals $\pm q$ such that $q$ occurs strictly
before $B_k$ in the enumeration $A_1,A_2,\ldots$. (Also note that
$C_N \notin cl(\mathcal R)$ because $C_N$ contains a splitting literal). 
All other leaf nodes
$N''$ of $\mathbb T_{T'}$ are also leaf nodes of $\mathbb T_T$ and we define
$C'_{N''} = C_{N''}$ and $\theta'_{N''} = \theta_{N''}$.
Then $(\mathbb T_{T'},C'_\bullet,\theta'_\bullet)$ is a failure witness for
$T'$ and we have $\mu^-(\mathbb T_{T'},C'_\bullet,\theta'_\bullet) <
\mu^-(\mathbb T_{T},C_\bullet,\theta_\bullet)$. 
Hence we have $\mu(\mathbb T_{T'},C'_\bullet,\theta'_\bullet) <
\mu(\mathbb T_T,C_\bullet,\theta_\bullet)$.
\qed
\endproof 

\begin{lemma}
If $T$ has a failure witness $(\mathbb T_T,C_\bullet,\theta_\bullet)$ 
and $T \predssplres T'$ then $T'\cup cl(\mathcal R)$ has a failure witness
$(\mathbb T_{T'\cup cl(\mathcal R)},C'_\bullet,\theta'_\bullet)$ with
$\mu(\mathbb T_{T'\cup cl(\mathcal R)},C'_\bullet,\theta'_\bullet) \le
\mu(\mathbb T_T,C_\bullet,\theta_\bullet)$.
\end{lemma}
\proof 
Let $C = C_1 \sqcup C_2 \in T$, $C_2$ is a non-empty $\mathbb P$-clause,
$C_1$ has at least one non-splitting literal, and
$T \predssplres T' = (T \setminus \{C\}) \cup \{C_1 \lor  - \ssp{C_2},
\ssp{C_2} \lor C_2\}$. 
If $C \neq C_N$ for any leaf node $N$ of $\mathbb T_T$ then there is nothing
to show.
Now suppose $C = C_N$ where $N$ is a leaf node of $\mathbb T_T$. If
$C_N \in cl(\mathcal R)$ then there is nothing to prove. Now suppose
$C_N \notin cl(\mathcal R)$.
As $C$ is constrained to contain at least one non-splitting literal, hence
the literal sequence $N$ has at least one non-splitting literal. By the
chosen enumeration $A_1,A_2,\ldots$, 
either $\ssp{C_2}$ or $-\ssp{C_2}$ occurs in the literal sequence $N$.
\begin{itemize}
\item If $\ssp{C_2}$ occurs in $N$ then $C_1 \lor  - \ssp{C_2}$ fails at
$N$. Let $N'$ be the highest ancestor of $N$ where it fails. 
$N'$ is a leaf node of $\mathbb T_{T'}$. We define
$C''_{N'} = C_1 \lor  - \ssp{C_2}$ and $\theta''_{N'} = \theta_N$. All other
leaf nodes $N''$ of $\mathbb T_{T'}$ are also leaf nodes of $\mathbb T_T$
and we define $C''_{N''} = C_{N''}$ and $\theta''_{N''} = \theta_{N''}$.
$(\mathbb T_{T'},C''_\bullet,\theta''_\bullet)$ is a failure witness
for $T'$. As $C_2$ has at least one non-splitting literal, we have
$\mu_1(C''_{N'},\theta''_{N'}) < \mu_1(C_N,\theta_N)$ 
(recall that $C_N\notin cl(\mathcal R)$) so that
$\mu(\mathbb T_{T'},C''_\bullet,\theta''_\bullet) \le
\mu(\mathbb T_T,C_\bullet,\theta_\bullet)$. As $T'\subseteq T'\cup
 cl(\mathcal R)$ hence the result follows.
\item
If $-\ssp{C_2}$ occurs in $N$ then $C_2 \lor \ssp{C_2}$ fails at
$N$. Since $C_1$ has at least one non-splitting literal, as in the previous
case, we obtain a failure witness $(\mathbb T_{T'},C''_\bullet,\theta''_\bullet)$
such that $\mu(\mathbb T_{T'},C''_\bullet,\theta''_\bullet) \le
\mu(\mathbb T_T,C_\bullet,\theta_\bullet)$.\qed
\end{itemize}
\endproof 

\begin{lemma}
If $T$ has a failure witness $(\mathbb T_T,C_\bullet,\theta_\bullet)$ 
and $T \splres T_1\mid T_2$ then $T_1\cup cl(\mathcal R)$ and
$T_2 \cup cl(\mathcal R)$ have failure witnesses
$(\mathbb T_{T_1\cup cl(\mathcal R)}\allowbreak ,\allowbreak  C'_\bullet\allowbreak ,\allowbreak  \theta'_\bullet\allowbreak )$ 
and $(\mathbb T_{T_2\cup cl(\mathcal R)},C''_\bullet,\theta''_\bullet)$ such that
$\mu(\mathbb T_{T_1 \cup cl(\mathcal R)},C'_\bullet,\theta'_\bullet) \le
\mu(\mathbb T_T,C_\bullet,\theta_\bullet)$
and $\mu(\mathbb T_{T_2 \cup cl(\mathcal R)},C''_\bullet,\theta''_\bullet) \allowbreak  \le\\
\allowbreak  \mu\allowbreak (\mathbb T_T, C_\bullet,\theta_\bullet)$.
\end{lemma}
\proof 
Let $C = C_1 \sqcup C_2 \in T$ such that $C_1$ and $C_2$ share no variables,
and we have $T \splres T_1 \mid T_2$ where $T_i = T \cup \{C_i\}$.
We prove the required result for $T_1$, the other part is symmetric.
If 
$C \neq C_N$ for any leaf node $N$ of $\mathbb T_T$ then there is nothing to 
show. Now suppose $C = C_N$ for some leaf node $N$ of $\mathbb T_T$.
If $C_N \in cl(\mathcal R)$  then there is nothing to show.
Now suppose $C_N \notin cl(\mathcal R)$. 
Since $C_1 \subseteq C$, hence $C_1$ also fails at $N$.
Let $N'$ be the highest ancestor of $N$ where $C_1$ fails. $N'$ is a leaf
node of $\mathbb T_{T_1}$. We define $C'''_{N'} = C$ and $\theta'''_{N'} = \theta$.
All other leaf nodes $N''$ of $\mathbb T_{T_1}$ are also leaf nodes of
$\mathbb T_T$, and we define $C'''_{N''} = C_{N''}$ and $\theta'''_{N''}
= \theta_{N''}$. $(\mathbb T_{T_1},C'''_\bullet,\theta'''_\bullet)$ is a failure
witness for $T_1$. Also $\mu_1(C'''_{N'},\theta'''_{N'}) \le 
\mu_1(C_N,\theta_N)$ (recall that $C_N \notin cl(\mathcal R)$). Hence
$\mu(\mathbb T_{T'},C'''_\bullet,\theta'''_\bullet) \le
\mu(\mathbb T_T,C_\bullet,\theta_\bullet)$. As $T_1\subseteq T_1\cup
cl(\mathcal R)$, hence the result follows.
\endproof

The following arguments are the ones that take care of replacement steps.
\begin{lemma}
If $T$ has a failure witness $(\mathbb T_T,C_\bullet,\theta_\bullet)$ 
and $T \replres T'$ then $T'\cup cl(\mathcal R)$ has a failure witness
$(\mathbb T_{T'\cup cl(\mathcal R)},C'_\bullet,\theta'_\bullet)$ with
$\mu(\mathbb T_{T'\cup cl(\mathcal R)},C'_\bullet,\theta'_\bullet) \le
\mu(\mathbb T_T,C_\bullet,\theta_\bullet)$.
\end{lemma}
\proof 
Let $C_1 = C'_1 \lor \pm A\sigma \in T$, $R = A \repl B \in \mathcal R$,
and $T \replres T' = (T \setminus \{C_1\})
\cup \{C\}$ where $C= C'_1 \lor \pm B\sigma$. 
If $C_1 \neq C_N$ for any leaf node of $\mathbb T_T$ then there is nothing to
prove. Now suppose that $C_1 = C_N$ for some leaf node $N$ of $\mathbb T_T$.
Let $N = \pm_1 A_1 \ldots \pm_k A_k$. If $C_1 \in cl(\mathcal R)$ then 
$T \subseteq T' \cup cl(\mathcal R)$, and there is nothing to prove. Now
suppose $C_1 \notin cl(\mathcal R)$.
We have a ground substitution $\theta$ such that
$C_1 \theta = C'_1\theta \lor \pm A\sigma\theta
\subseteq \{\mp_1 A_1, \ldots, \mp_k A_k\}$.
As $R$ is ordered we have $A \atge B$. Hence 
$A\sigma\theta \atge B\sigma\theta$.
Hence  either
$\pm B\sigma\theta \in \{\mp_1 A_1, \ldots, \mp_k A_k\}$ or
$\mp B\sigma\theta \in \{\mp_1 A_1, \ldots, \mp_k A_k\}$. 
\begin{itemize}
\item
Suppose $\pm B\sigma\theta \in \{\mp_1 A_1, \ldots, \mp_k A_k\}$.
Since $C_1 \theta = C'_1\theta \lor \pm A\sigma\theta
\subseteq \{\allowbreak \mp_1 \allowbreak A_1, \allowbreak \ldots,\allowbreak  \mp_k\allowbreak  A_k\allowbreak \}$, hence
$C \theta = C'_1\theta \lor \pm B\sigma\theta
\subseteq \{\mp_1 A_1, \ldots, \mp_k A_k\}$. Hence $C$ fails at $N$.
Let $N'$ be the highest ancestor of $N$ where $C$ fails. $N'$ is a leaf
node of $\mathbb T_{T'}$. We define $C''_{N'} = C$ and $\theta''_{N'} = \theta$.
All other leaf nodes $N''$ of $\mathbb T_{T'}$ are also leaf nodes of
$\mathbb T_T$, and we define $C''_{N''} = C_{N''}$ and $\theta''_{N''}
= \theta_{N''}$. $(\mathbb T_{T'},C''_\bullet,\theta''_\bullet)$ is a failure
witness for $T'$. Also $\mu_1(C''_{N'},\theta''_{N'}) \le 
\mu_1(C_N,\theta_N)$ (recall that $C_N \notin cl(\mathcal R)$). Hence
$\mu(\mathbb T_{T'},C'_\bullet,\theta'_\bullet) \le
\mu(\mathbb T_T,C_\bullet,\theta_\bullet)$. As $T'\subseteq T'\cup
cl(\mathcal R)$, hence the result follows.
\item Suppose $\mp B\sigma\theta \in \{\mp_1 A_1, \ldots, \mp_k A_k\}$.
Since $\pm A\sigma\theta = \in 
\{\mp_1 A_1, \ldots, \mp_k A_k\}$, hence the clause $\mp A \lor \pm B 
\in cl(\mathcal R)$ fails at $N$. 
Let $N'$ be the highest ancestor of $N$ where $\mp A \lor \pm B$ fails.
$N'$ is a leaf node of $\mathbb T_{T'\cup\{\mp A \lor \pm B\}}$.
We define $C''_{N'} = C$ and $\theta''_{N'} = \theta$.
All other leaf nodes $N''$ of $\mathbb T_{T'\cup\{\mp A \lor \pm B\}}$
are also leaf nodes of
$\mathbb T_T$, and we define $C''_{N''} = C_{N''}$ and $\theta''_{N''}
= \theta_{N''}$. $(\mathbb T_{T'\cup\{\mp A \lor \pm B\}},C''_\bullet,
\theta''_\bullet)$ is a failure
witness for $T'\cup\{\mp A \lor \pm B\}$. Also $\mu_1(C''_{N'},\theta''_{N'}) \le
\mu_1(C_N,\theta_N)$ since $\mu_1(C''_{N'},\theta''_{N'})$ is the empty
multiset. Hence
$\mu(\mathbb T_{T'\cup\{\mp A \lor \pm B\}},C''_\bullet,\theta''_\bullet) \le
\mu(\mathbb T_T,C_\bullet,\theta_\bullet)$. As $T'\cup\{\mp A \lor \pm B\}
\subseteq T'\cup
cl(\mathcal R)$, hence the result follows.
\qed
\end{itemize}
\endproof 

For a tableaux $\mathcal T = S_1 \mid \ldots \mid S_n$, define
$\mathcal T \cup S = S_1 \cup S \mid \ldots \mid S_n \cup S$. We define
a failure witness for such a $\mathcal T$ to be a multiset
$\{(\mathbb T_{S_1},C^1_\bullet,\theta^1_\bullet),\ldots,
(\mathbb T_{S_n},C^1_\bullet,\theta^n_\bullet)\}$ where
each $(\mathbb T_{S_i},C^i_\bullet,\theta^i_\bullet)$ is a failure
witness of $S_i$. We define\\ 
\strut \hfill $\mu(\{\mathbb T_{S_1},C^1_\bullet,\theta^1_\bullet),\ldots,
(\mathbb T_{S_n},C^1_\bullet,\theta^n_\bullet\}) = 
\{\mu(\mathbb T_{S_1},C^1_\bullet,\theta^1_\bullet),\ldots,
\mu(\mathbb T_{S_n},C^1_\bullet,\theta^n_\bullet)\}$.\hfill\ \\
Then it is clear that $S\cup cl(\mathcal R)$ 
has a failure witness and whenever any $\mathcal T$
has a failure witness in which one of the trees has at least two nodes,
then $\mathcal T \atordselfssplreplres \mathcal T'$ for some 
$\mathcal T'$ such that
$\mathcal T'\cup cl(\mathcal R)$ has a strictly smaller failure witness.
Hence we have some $\mathcal T$ such that
$S \atordselfssplreplres^* \mathcal T$ and $\mathcal T\cup cl(\mathcal R)$ 
has a failure witness
in which each tree is a root node. Then $\mathcal T \cup cl(\mathcal R)$
is closed. Hence $\mathcal T$ is closed.
\qed
\endproof 
\restorethcounter


\end{document}